\documentclass[pdflatex,sn-basic]{sn-jnl}

\usepackage{graphicx}%
\usepackage{multirow}%
\usepackage{amsmath,amssymb,amsfonts}%
\usepackage{amsthm}%
\usepackage{mathrsfs}%
\usepackage[title]{appendix}%
\usepackage{xcolor}%
\usepackage{textcomp}%
\usepackage{manyfoot}%
\usepackage{booktabs}%
\usepackage{listings}%
\usepackage{multirow}
\usepackage{mathtools}
\graphicspath{{figs/}}

\begin{document}

\title[Article Title]{Water versus land on temperate rocky planets}


\author*[1]{\fnm{Claire Marie} \sur{Guimond}}\email{claire.guimond@physics.ox.ac.uk}

\author[2]{\fnm{Tilman} \sur{Spohn}}

\author[3]{\fnm{Svetlana} \sur{Berdyugina}}

\author[4]{\fnm{Paul K.} \sur{Byrne}}

\author[5]{\fnm{Nicolas} \sur{Coltice}}

\author[6, 7]{\fnm{Donald M.} \sur{Glaser}}

\author[8]{\fnm{Manasvi} \sur{Lingam}}

\author[9]{\fnm{Charles H.} \sur{Lineweaver}}

\author[10]{\fnm{Peter A.} \sur{Cawood}}

\affil*[1]{\orgdiv{Atmospheric, Oceanic, and Planetary Physics}, \orgname{University of Oxford}, \orgaddress{\street{Parks Rd}, \city{Oxford} \postcode{OX1 3PU}, \country{United Kingdom}}}

\affil[2]{\orgdiv{Institute of Space Research}, \orgname{German Aerospace Center (DLR)}, \orgaddress{\street{Rutherfordstr. 2}, \city{Berlin}, \postcode{12489}, \country{Germany}}}

\affil[3]{\orgdiv{Istituto ricerche solari Aldo e Cele Dacc\'o (IRSOL), Faculty of Informatics}, \orgname{Universita della Svizzera italiana}, \orgaddress{\street{Via Patocchi 57}, \city{Locarno}, \postcode{6605}, \country{Switzerland}}}

\affil[4]{\orgdiv{Department of Earth, Environmental, and Planetary Sciences}, \orgname{Washington University in St. Louis}, \orgaddress{\street{1 Brookings Drive}, \city{St. Louis}, \state{MO}, \postcode{63130}, \country{USA}}}

\affil[5]{\orgdiv{UMR GéoAzur}, \orgname{Université Côte d'Azur}, \orgaddress{\street{250 rue Albert Einstein}, \city{Valbonne}, \country{France}}}

\affil[6]{\orgname{Goddard Institute for Space Studies}, \orgaddress{\street{2880 Broadway}, \city{New York}, \state{NY}, \postcode{10025}, \country{USA}}}

\affil[7]{\orgname{Blue Marble Space Institute of Science}, \orgaddress{\street{600 1st Avenue}, \city{Seattle}, \state{WA}, \postcode{98104}, \country{USA}}}

\affil[8]{\orgdiv{Department of Aerospace, Physics and Space Sciences}, \orgname{Florida Institute of Technology}, \orgaddress{\street{150 W. University Blvd.}, \city{Melbourne, FL}, \postcode{32901}, \country{USA}}}

\affil[9]{\orgdiv{Research School of Astronomy \& Astrophysics}, \orgname{Australian National University}, \orgaddress{\city{Canberra}, \postcode{2600}, \country{Australia}}}

\affil[10]{\orgdiv{School of Earth, Atmosphere \& Environment}, \orgname{Monash University}, \orgaddress{\street{9 Rainforest Walk}, \city{Clayton, Victoria}, \postcode{3800}, \country{Australia}}}

\abstract{
Water and land surfaces on a planet interact in particular ways with gases in the atmosphere and with radiation from the star. These interactions define the environments that prevail on the planet, some of which may be more amenable to prebiotic chemistry, some to the evolution of more complex life. This review article covers (i) the physical conditions that determine the ratio of land to sea on a rocky planet, (ii) how this ratio would affect climatic and biologic processes, and (iii) whether future astronomical observations might constrain this ratio on exoplanets. Water can be delivered in multiple ways to a growing rocky planet --- and although we may not agree on the contribution of different mechanism(s) to Earth's bulk water, hydrated building blocks and nebular ingassing could at least in principle supply several oceans' worth. The water that planets can sequester over eons in their solid deep mantles is limited by the water concentration at water saturation of nominally anhydrous mantle minerals, being in sum likely less than 2000 ppm of the planet mass. Water is cycled between mantle and surface through outgassing and ingassing mechanisms that, while tightly linked to tectonics, do not necessarily require plate tectonics in every case. The actual water/land ratio at a given time then emerges from the balance between the volume of surface water on the one hand, and on the other hand, the shape of the planet (its ocean basin volume) that is carved out by dynamic topography, the petrologic evolution of continents, impact cratering, and other surface-sculpting processes. By leveraging the contrast in reflectance properties of water and land surfaces, spatially resolved 2D maps of Earth-as-an-exoplanet have been retrieved from models using real Earth observations, demonstrating that water/land ratios of rocky exoplanets may be determined from data delivered by large-aperture, high-contrast imaging telescopes in the future.
}


\maketitle

\section{Introduction}

\begin{figure}
    \centering
    \includegraphics[width=1\linewidth]{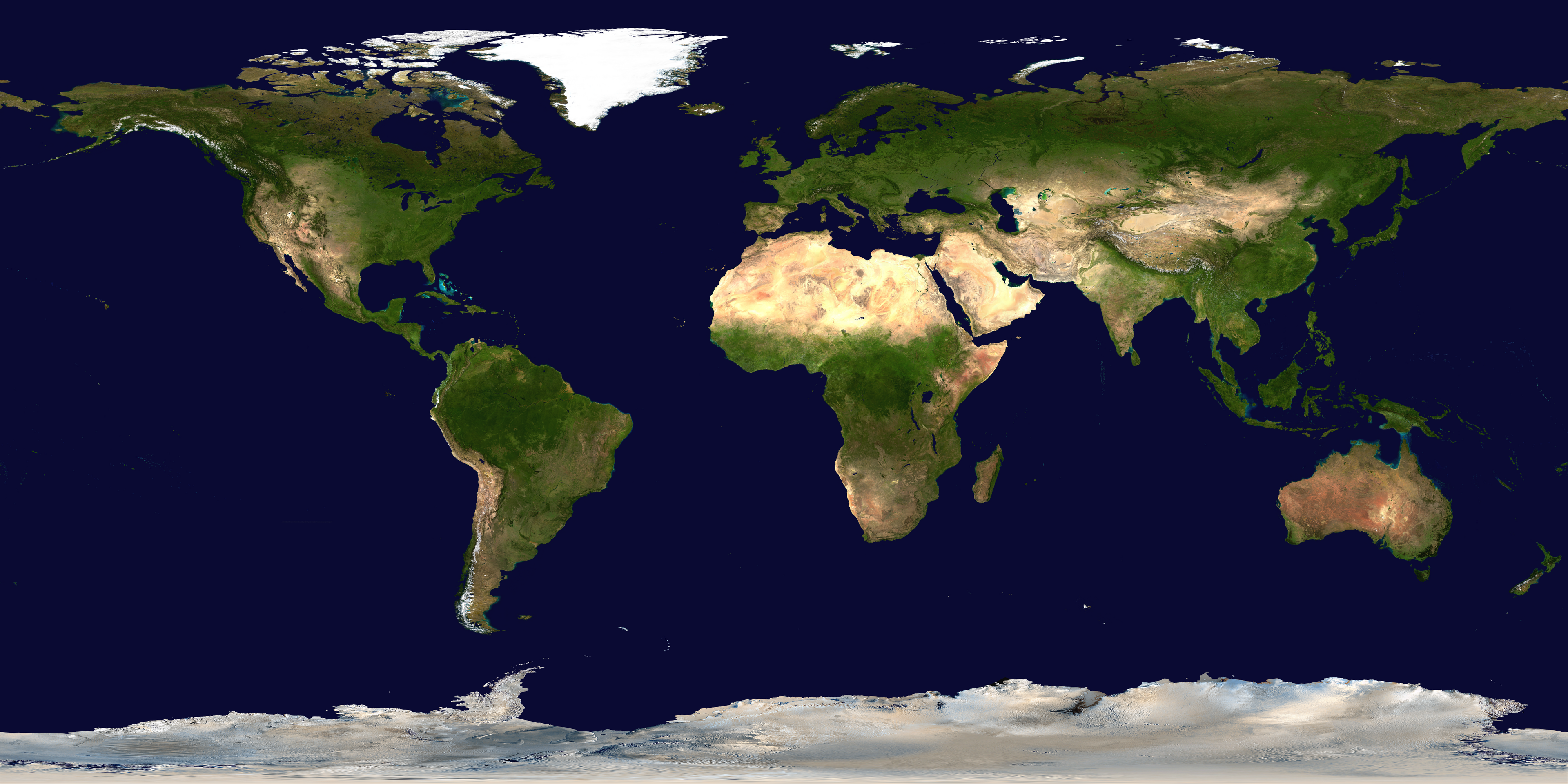}
    \caption{Composite image of the Earth's surface compiled from NASA’s Terra satellite MODIS images, with clouds removed. Credit: NASA Goddard Space Flight Center Image by Reto Stöckli (land surface, shallow water, clouds). Enhancements by Robert Simmon (ocean color, compositing, 3D globes, animation). Data and technical support: MODIS Land Group; MODIS Science Data Support Team; MODIS Atmosphere Group; MODIS Ocean Group, Public Domain, https://commons.wikimedia.org/w/index.php?curid=50497070}
    \label{fig:blue marble}
\end{figure}

The world map of Earth is two thirds oceanic-blue and one third continental-brown (Fig. \ref{fig:blue marble}). For most of us, this map is our mental image, for better or worse, of a habitable planet. Yet studying Earth history has not clearly answered whether this condition of contemporaneous water and land has been a prerequisite for our planet's long-term state of hosting life. The existence of billions of distant exoplanets in our Galaxy could, in principle, serve as a means to overcome this anthropic bias, although observing even a few of them in the necessary detail is one of the field's multigenerational challenges. In the meantime, the possible existence of so many rocky planets with different geological histories and bulk properties than Earth has motivated theoretical research into why --- and to what extent --- some planets might sustain oceans next to land, and how this ocean coverage would affect climates, biospheres, and their detectable signatures.

Working to understanding the drivers and consequences of planet surface character is in line with recent discussions of planetary `habitability', going beyond stellar luminosity and atmospheric chemistry to include other planetary properties: the presence of a magnetic field, the diversity of geological provinces, and geodynamic processes such as plate tectonics that drive the interior heat engine and deep volatile cycling \citep[e.g.,][and see Glaser et al., 2026; Spohn et al., 2026, this topical collection]{lineweaver2012,heller_superhabitable_2014,schulze-makuch_superhabitable_2020, scherf_eta-earth_2024, stern_2024}. In the present article, we focus on the importance of ocean- versus continent-covered surface area, hereafter called the water/land ratio.


Agnostic to any future extraterrestrial life detection, such processes are also interesting for understanding how terrestrial planets evolve over time. Why does Earth have just enough liquid water on its surface to cover some, but not all, of its land? If some process(es) keep(s) Earth's oceans from either flooding the land or disappearing over billions of years, could the same process(es) be operating on other rocky planets? How do planets sculpt their crusts into continents and mountains, raising land above sea level? It is interesting to note that water, playing a fundamental role for life, has a similar fundamental role in plate tectonics. Based on their study of the evolution of continents, \cite{campbell_water_1983} wrote `no water, no granites --- no oceans, no continents' or as \cite{karato_water_2015} put it `no water, no oceans, no plate tectonics, no granites, no continents' --- to which one uncertainly might add, `no life?'. Therefore, we ask: how does the distribution of land and oceans on a planet control its climate, and its potential to host life? Might life itself act to help regulate water cycling \citep{harding2009}?  Could future astronomical observations identify water or land areas and map the surfaces of exoplanets? Although speculative, these questions are already being approached via multiple, interdisciplinary angles. 

Here we consolidate the knowledge needed to start answering these questions by reviewing the growing body of work on the spatial extents of water versus land on rocky planets. First, Section \ref{sec:topography} reviews how processes building topography carve out ocean basins, and how these mechanisms could operate across different rocky planets. Section \ref{sec:water-budget} then provides an overview of how water is delivered to and partitioned between reservoirs inside and on the surface of the planet. Section \ref{sec:consequences} discusses some consequences of water-versus-land ratios on planetary climate and potential biospheres. Finally, Section \ref{sec:EPSI} addresses how to detect and map water and/or land on exoplanets using future observatories.

\section{Desert, ocean, or blue marble: what sets a planet's water/land ratio?}


At any point in time, a planet's water/land ratio represents the trade-off between \textit{(i)} the volumetric capacity of its ocean basins \citep[e.g.,][]{honing_biotic_2014, cowan_water_2014, honing_continental_2016, simpson_bayesian_2017, honing_land_2023, dong_constraining_2021, guimond_blue_2022}, and \textit{(ii)} the actual volume of surface water. The volume of surface water depends on \textit{(ii.A)} the total bulk inventory of water available, and \textit{(ii.B)} how this water is distributed between planetary interior and surface. The volumetric capacity of the ocean basins depends on the height and distribution of topography. Section \ref{sec:topography} explores point \textit{i}; Section \ref{sec:water-budget} explores point \textit{ii}.

In sum, a rocky planet's bulk water inventory is largely set during its formation (Section \ref{sec:water-delivery}), but will decrease with time due to atmospheric loss to space (see Kubyshkina et al., 2026 ,this topical collection). This water budget will be distributed between different reservoirs throughout the planet, the sizes of which themselves have physical limits (Section \ref{sec:deep-water-cycle}). The amount of water stored in the solid mantle cannot exceed the water solubility limits of its constituent minerals \citep[e.g.,][]{karato_water_2015, dong_constraining_2021, shah_internal_2021, guimond_mantle_2023}, which are orders of magnitude lower than the solubility of water in magma \citep[e.g.,][]{sossi_solubility_2023}. Some of the planet's water, if accreted before the segregation of its iron core, will partition into this metal and be locked there forever. At the surface,\footnote{There is an additional reservoir for water at a planet's (sub)surface: groundwater, which we neglect given that it represents $\sim$6 ppm of the ocean mass on Earth. Frozen surface water (glaciers and ice caps) could lock in larger fractions of the hydrosphere, and is discussed in the context of water/land ratio effects on climate in Section \ref{sec:climate-albedo}.} the capacity of ocean basins to `store' water is further limited by the shape of the planet's rocky surface---the volume of empty space contained below its highest point (Section \ref{sec:topography}). Excess water beyond this volume would inundate the planet's entire land mass.

Over time, the ocean volume can change according to the fluxes of water from the mantle to the surface (\textit{outgassing}) and vice versa (\textit{ingassing}) that control how fully each reservoir for water is filled (Section \ref{sec:deep-water-cycle}). The term `(volcanic) outgassing', as used in this context, condenses a chain of processes. First, the convecting, volatile-bearing mantle adiabatically decompresses to create melt. Second, this melt migrates through the solid silicate interior to the surface. Third, volatiles exsolve from this melt at lower surface pressures, where they are less soluble, to ultimately produce volcanic gas. Meanwhile, ingassing can refer to a number of processes that act to bury hydrated, solid crust into the mantle. Rates of ingassing and outgassing are therefore highly dependent on the tectonic mode\footnote{The term \textit{tectonic mode} encompasses the overall dynamical nature of the `couplings' between the surface and interior \citep[see][]{lenardic_diversity_2018, stern_tectonic_2023}.} of the planet \citep[e.g.,][and see Lourenco et al., 2026, this topical collection]{kite_geodynamics_2009, noack_can_2014, komacek_effect_2016, foley_whole_2016, seales_deep_2020, spaargaren_influence_2020}, as well as the potential for minerals in the crust to themselves become hydrated \citep[e.g., higher on Mars than on Earth;][]{wade_divergent_2017, scheller_longterm_2021}.

\subsection{Planetary topography and the capacity of ocean basins}\label{sec:topography}

\subsubsection{The shape of an active planet: Dynamic topography controlled by mantle convection}\label{subsec:dynamic}


\begin{figure}
    \centering
    \includegraphics[width=0.75\linewidth]{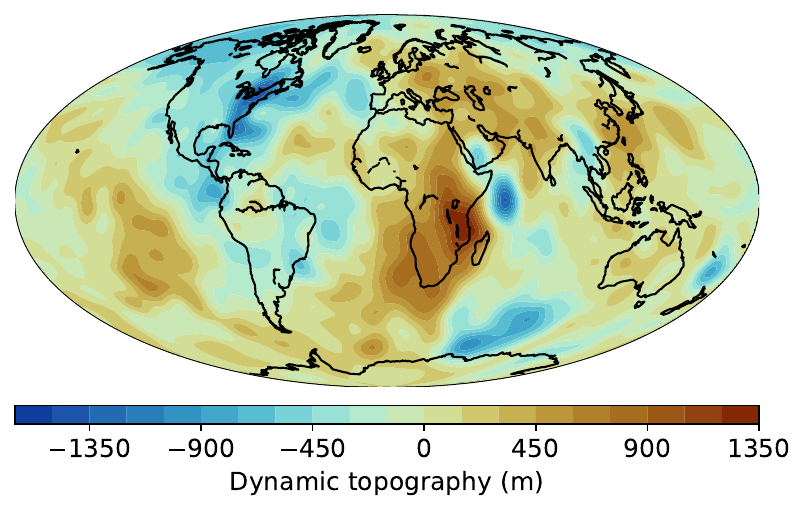}
    \caption{The dynamically-supported component of Earth's topography as calculated by \citet{straume_impact_2024}, based on the SMEAN2 seismic tomography model. Data are available at \url{https://doi.org/10.5281/zenodo.8262689}. Note topography supported by compositional isostasy (e.g., continents), which is excluded from this map, makes up an additional component to the total topography.
    \label{fig:dt}}
\end{figure}

Topography on a dynamic planet has two primary components. The isostatic component mainly reflects compositional (density) differences within the crust\footnote{In the Pratt model; the Airy isostasy model assumes constant crustal density but variable crustal thickness.} and is often conceptualised as rigid columns of rock extending above a compensation depth. This isostatic balance explains major topographic contrasts on Earth, such as the elevation differences between continents and ocean basins (see Section \ref{subsec:continents}), and has been studied extensively for over a century. The rheological concepts of the lithosphere and asthenosphere were developed in part to explore this phenomenon \citep{barrell1914}. 

The other component arises from mantle convection and is referred to as dynamic topography \citep{pekeris1935}, with its global average being zero (Fig. \ref{fig:dt}). A simplistic version of dynamic topography holds that it consists of topographic highs above hot (upwelling) mantle convection currents, and lows for cold (downwelling) currents. Buoyant anomalies generate upwellings and dynamic topographic highs. For instance, the African superswell on Earth, associated with mantle plumes, corresponds to a dynamic topographic high of $>$500\,m \citep{lithgow1998}, whereas the margins of the Pacific, near subduction trenches, are linked to dynamic topographic lows of about 2 km \citep{ricard1993}. 

On Earth, characterising dynamic topography presents a major observational challenge because the isostatic component must be `removed' to reveal it (see Fig. \ref{fig:dt}). The task of measuring Earth's dynamic topography is ongoing, with density models for both oceanic and continental lithospheres being developed to differentiate between the two \citep{hoggard2021}. The challenge is even greater for Venus, where topographic observations are less precise and the tectonic mode responsible for generating internal density heterogeneities remains uncertain. However, based on long-wavelength deformation in the Baltis Vallis region, \citet{mcgregor_probing_2025} suggested that dynamic topography on Venus may be lower than on Earth.  

Vital clues to the origins and support mechanism of topography come in comparing spherical harmonic power spectra. Isostatic and dynamic topography generate distinct geoid signatures. On Earth, isostatically supported topography correlates well with the geoid at spherical harmonic degrees greater than 10, corresponding to shorter wavelengths \citep{lestunff1995}. In contrast, long-wavelength topography, particularly at degree 2 (corresponding to a wavelength of $\sim$20,000\:km), is generally interpreted as the surface expression of dynamic topography driven by mantle convection \citep{steinberger2019}. Although the total topography shows poor correlation with the geoid at these long wavelengths, the predicted dynamic topography alone does correlate  \citep{Flament2019}. Notably, the spherical harmonic power spectrum of seismic anomalies in the lower mantle is also dominated by degree 2, supporting the interpretation that these anomalies reflect deep-seated density variations and associated mantle flow. Contributions at other low degrees are typically attributed to density anomalies and flow within the upper mantle \citep{ricard1993}.


This observational challenge has driven the development of theoretical predictions for dynamic topography. Predictions of Earth's dynamic topography have been a subject of extensive debate and modelling \citep{flament2013, davies2023}. Since dynamic topography is a consequence of mantle flow, it reflects the interaction between buoyancy variations and the rheology of the deep interior. Predictive dynamic topography models are thus ultimately based on mantle buoyancy variations and rheology. The viscosity structure of the mantle influences the magnitude of dynamic topography because this structure governs how stress is distributed to deform the interfaces \citep{flament2013}. 

Both buoyancy heterogeneity and 3D viscosity result from self-organization during mantle convection. As convective vigour increases, thermal mixing intensifies, reducing the size and amplitude of thermal heterogeneities. 
Convective vigour and rheology are tightly interwoven with the emergent tectonic mode of a planet \citep[see also Lourenco et al., 2026, this topical collection]{solomatov_parameterization_1993, lenardic_diversity_2018}, so a key consideration in dynamic topography is the evolution of tectonic modes over time. Looking at models, however, some general patterns have become clear across the spectrum of tectonic modes:
\begin{enumerate}
    \item If viscosity were constant throughout a planetary mantle, increasing convective vigour reduces the amplitude of dynamic topography \citep{bercovici1988}, but leads to more rapid fluctuations. In the extreme case, as convective vigour approaches infinity, topography approaches zero. The power spectrum of dynamic topography directly reflects the dominant wavelengths of mantle flow. 
    \item If a planet's viscosity increases strongly with decreasing temperature, its (cold) surface may form a stagnant lid; the surface is locally too viscous to participate in convection. In such cases, plumes and small-scale instabilities beneath the viscous lid generate topography, whereas most of the mantle interior remains homogeneous. The buoyancy sources for dynamic topography are limited, and their surface expressions are filtered by the high viscosity contrast between the lid and the mantle. On a stagnant lid planet, dynamic topography amplitude still decreases with --- but depends more weakly on --- convective vigour, as long as there is this large viscosity contrast between the lid and mantle \citep{guimond_blue_2022, mcgregor_probing_2025}. Shorter wavelengths become more prominent in the power spectrum. 
    \item If viscosity also depends on stress, the planet may under some conditions enter the plate tectonics mode, with persistent subduction \citep{tackley2023}. In this case, again the amplitude of dynamic topography decreases with increasing convective vigour. With variable viscosity, intermediate wavelengths in the power spectrum are more pronounced than with uniform viscosity \citep{arnould2018}. Considering the possibility of changing tectonic modes over time, transient or extinct subduction could have left cold slabs sinking into the mantle, driving deep mantle flow and creating long-wavelength topographic lows.
    \item Under scenarios involving intense melting (e.g., young and/or strongly heated planets), the planet may be described by a heat-pipe or squishy-lid tectonic mode \citep{moore2013, lourenco2020}. Here, magma intrudes in the lithosphere and stalls, creating strong density variations within the lithosphere. Heat-pipe tectonics are associated with an extreme viscosity contrast between a partially molten mantle and a strong lid, resulting in dynamic topography similar to that of a stagnant lid regime. In the squishy-lid regime, the lid is softer, leading to larger dynamic topography amplitudes, particularly localized around plumes. Squishy-lid dynamics also generate transient downwellings, which could produce stronger long-wavelength components, such as those generated by deep subducting slabs today.
\end{enumerate}
In summary, more-vigorous convection --- i.e., at high Rayleigh number --- is generally associated with lower dynamic topography amplitudes. The Rayleigh number increases with larger mantles, hotter temperatures, and lower viscosities (due themselves to hot temperatures, or high mantle water content), which for all other things being equal are expected for more massive rocky planets. Smaller, older, and colder planets with lower convective vigour should exhibit larger topographic amplitudes, influenced by the viscosity contrast between the lithosphere and mantle \citep{guimond_blue_2022}. Meanwhile, the power spectrum of dynamic topography --- its spatial distribution --- is closely tied to the planet’s tectonic mode. High amplitudes at low degrees are expected on planets with deep recycling of lithosphere, such as Earth’s plate tectonics, with stagnant lid and squishy lid/heat pipe regimes showing shorter-wavelength dynamic topography.

With the above said, however, the topography of Mars shows that enigmatic patterns can emerge even on stagnant lid planets, due to other topographic support mechanisms beyond dynamic topography. Mars' long-wavelength topography is dominated by the isostatic component \citep{neumann_crustal_2004}. Degree 1 corresponds to the crustal dichotomy: half of Mars is thick crust in the southern hemisphere; and the other half thinner crust in the northern hemisphere. Degree 2 corresponds to Tharsis (thick crust) and Hellas annulus (thin crust). Nonetheless, there is no obvious long-wavelength dynamic topography observed on Mars. 


How is a planet's water/land ratio related to its dynamic topography? For most planets, high (dynamic) topography is expected to decrease the water/land ratio by creating elevation contrasts. However, if dynamic topographic highs occur beneath isostatic topographic lows, and vice versa, the water/land ratio may increase. This situation may arise if cold convection currents (or cold structures such as slabs) occur beneath thick, buoyant lithosphere. If positive dynamic topography is concentrated in oceanic regions, the water/land ratio is maximized. Under typical scenarios, however, thick lithosphere acts as a conductive lid beneath which warmer mantle rises, either as plumes \citep{guillou1995} or diffuse upwellings \citep{phillips2010}. The actual sea level by definition follows the geoid (equal gravity-field potential): sea-level variations are expressed by the difference between the geoid and dynamic topography over continent-like areas. 
Nonetheless, the 3D model of \citet{schreiber_maia_viscosity_2024} shows that geoid height fluctuations are small: consistently no more than 10\% of the dynamic topography, across a range of mantle viscosity structures for a Venus-size planet (J. Schreiber Maia, personal communication). Hence we can view dynamic topography as a fundamental and intrinsic source of ocean basin capacity variation on geodynamically-active rocky planets.


\subsubsection{Higher-order topography: The evolution of continents}\label{subsec:continents}


\begin{figure}
    \centering  
         \includegraphics[width=0.9\linewidth]{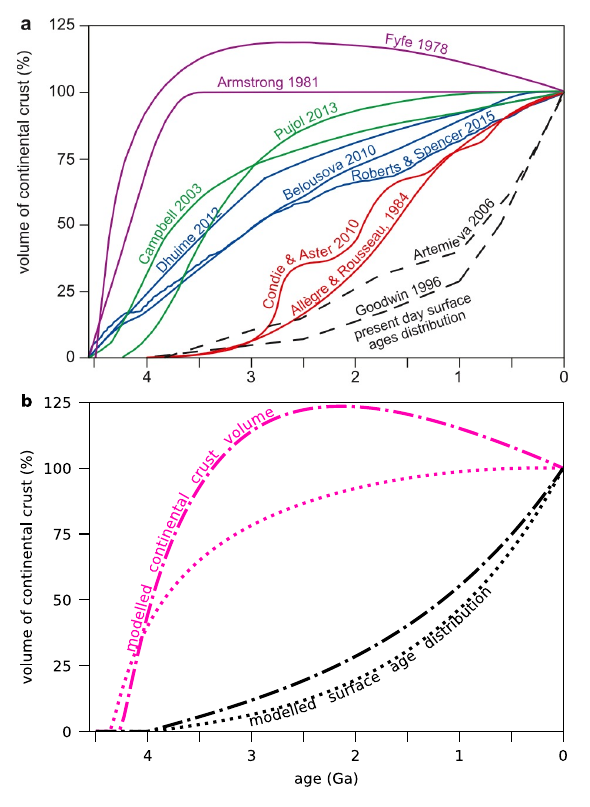}
   \caption{\textit{(a):} Models of the growth of the continental crustal volume, normalised to the present-day continental crust volume, compared with the present-day cumulative surface age distribution of \cite{goodwin_precambrian_1996}. Figure adapted from \cite{cawood_secular_2022}. Note that a few more recent curves \citep{rosas_rapid_2018,guo2020} follow \citet{armstrong_radiogenic_1981}. \textit{(b):} Continental crust growth from two models of \cite{honing_land_2023} (pink lines), compared with the same models' cumulative age distributions at the present day (black lines). Dash-dotted lines show models where 2/3 of the crustal production rate is proportional to the mantle flow velocity and 1/3 is proportional to the continent surface area; dotted lines show the reverse. All models consider thermal blanketing by continents and partitioning of heat-producing radiogenic elements into the crust. 
   The cumulative surface age distributions are similar despite the substantially different growth curves, due to continuous production and recycling of the crust \citep[see also][]{korenaga2021land}.
   }
    \label{fig:Crust Age}
\end{figure}

Modern Earth's markedly bimodal distribution of surface elevation comes from the fact that its continents are made from lower density rock than its oceanic crust. Buoyant felsic\footnote{The term `felsic' is a portmanteau of the mineral names feldspar and silica (i.e., quartz), and refers to rocks rich in such minerals. `Mafic' is likewise a portmanteau of `magnesium' and 'ferric'.} rock, chiefly granite, dominates the continents; denser mafic rock, mainly basalt, composes the oceanic crust. Oceanic crust is produced by partial decompression melting\footnote{In partial melting of a source rock, the minerals with the lowest melting temperatures are extracted to make a new rock, leaving the minerals with the highest melting temperatures behind at the source. A reliable way to bring mantle rock above its melting temperature is adiabatic decompression upon upwelling.} of mantle rock beneath mid-oceanic ridges. It is returned to the mantle in subduction zones.  Continental crust is produced at subduction zones (in admittedly very small net amounts in the modern era), for subduction of water, which depresses the solidus temperature, promotes partial melting of oceanic crust along with subducted sediments and mantle rock \citep[e.g.,][]{campbell_water_1983, stein_mechanisms_2015}. If the continents were more mafic and closer to oceanic crust in density, then surface elevations would be unimodally distributed. Indeed, early Archean crust before $\sim$3.5 Ga is thought to have been more mafic, with relief correspondingly flatter \citep{dhuime_emergence_2015, tang_archean_2016, cawood_continental_2019, chen_how_2020, palin_mafic_2021}.


Although high continental relief appears to be, unlike dynamic topography, unique to Earth, the presence of these buoyant continents is part of the key to why Earth sustains a stable intermediate water/land ratio. Some have speculated about felsic highlands on Venus \citep[e.g.,][as discussed below]{hashimoto_felsic_2008, gilmore_virtis_2015}, for which planned Venus missions may provide substantiating data. Arguments have also been made for granitic crust forming early on Mars \citep{mcsweenjr_chemical_1999, bandfield_identification_2004, michalski_diverse_2024, lee_crustal_2025, malarewicz_evidence_2025}, as have arguments for later silica-rich igneous crust \citep{czarnecki_identification_2020}. Apart from these still-mysterious examples, the lack of clear evidence for large-scale granitic continents beyond Earth makes it hard to predict how often we can expect them to show up elsewhere. 

Still, having an understanding of why Earth's modern continental crust is on average 35--40 km thick (versus 6--7 km for oceanic crust), with a felsic composition, will give us the necessary context to thinking about any analogous hypsometric phenomenon on other planets. The detailed timing and provenance of the continents is still hotly debated \citep[Fig. \ref{fig:Crust Age}a; see][for a review]{chowdhury_continental_2024}. It is generally agreed that Earth started as a water planet \citep[e.g.,][]{campbell_water_1983, karato_water_2015, chowdhury_continental_2024}. 
Elevated continental land masses then evidently were seeded, rose from the seas, and grew in area over time, such that continental crust spans about 40\% of Earth's surface today. Of this percentage, 87\% is emerged; 13\% is continental shelf covered by water. In total, 29\% of the Earth's surface is subaerial land.  \cite{chowdhury_continental_2024} recently reviewed the substantial evidence for subaerial land surfaces by 3.5 Ga, although these surfaces may date back even to the Hadean \citep[e.g.,][]{oneil_earth_2024}. Observational constraints on the area of Earth's continental crust, and more generally on the area of its emerged land surfaces, come from the age distribution of continental crust rock (Fig. \ref{fig:Crust Age}), the freeboard as a function of time, and geological and geochemical evidence \citep[e.g.,][]{cawood_secular_2022,chowdhury_continental_2024}. 

Multiple lines of evidence reviewed in \citet{cawood_continental_2019} point to a change towards roughly modern hypsometry on Earth circa 3 Ga. The initially mafic and unimodal Archean crust is thicker than modern oceanic crust because of the hotter mantle temperature and associated degree of melting. Melting of metamorphosed mafic rock at the base of the crust (e.g., amphibolite) could have produced the granitic rock characteristic to the Archean (tonalite–trondhjemite–granodiorite or TTG). The crust first begins to increase in both thickness and overall silica content with the emplacement of TTGs: the increasing variability of certain geochemical tracers in the rock record around 3 Ga suggests a corresponding deviation from unimodality in crust thickness and hence hypsometry \citep{moyen_archaean_2018}. With the transition to plate tectonics, melting at convergent oceanic plate boundaries started to produce new crust with an average composition of (felsic) calc-alkaline andesite \citep{rudnick_composition_2003}, possibly establishing the thick continents in the long term. 

Although associating the timing of modern hypsometry with the timing of plate tectonics is certainly tempting, debate about the connections between plate tectonics initiation and continental evolution persists in part because the rock record becomes increasingly sparse going back in time. It is widely accepted that plate tectonics has been operating through the Proterozoic, since about 2.5 Ga at the latest. Yet the beginnings of emerged land and proto-continents would have been seeded in the Archean \citep[e.g.,][]{cawood_continental_2013, cawood_secular_2022,chowdhury_continental_2024, frost_cratons_2024}.

The advent of plate tectonics may have also stabilised Earth's continental area due to several potential feedback mechanisms: between plate boundary type and new crust composition and/or between continental area and lithospheric stress, for example \citep[e.g.,][]{Lenardic:2005, honing_continental_2016, cawood_continental_2019}. The observed cumulative age distribution of continental crust (Fig. \ref{fig:Crust Age}b) can be reproduced in models which assume some form of continuous continental crust production and recycling occurring through the Archean, or even the Hadean. \citet{honing_land_2023} modelled the crust production rate as proportional to the vigour of mantle convection, and the recycling rate as proportional to the area of continental crust. Under this formulation, continental growth rates would decrease with time as the mantle cools and convects less vigorously, whereas recycling rates would increase with time: rapid early crust formation that turns into a steady-state \citep[see Fig. \ref{fig:Crust Age}b]{fyfe_evolution_1978, armstrong_radiogenic_1981, rosas_rapid_2018, guo2020}. Yet different assumptions about the extents of these proportionalities can result in the same cumulative growth curve, despite the actual areal growth curves differing considerably (Fig. \ref{fig:Crust Age}b). In sum --- and as reviewed in \citet{korenaga2021land} --- definitive interpretations of the evidence remain elusive.

So is Earth's continental surface area of 40\% a determined outcome of plate tectonics? \citet{simpson_bayesian_2017} used Bayesian statistics to argue that Earth should be on the dry tail of a mostly water-rich distribution. The statistical argument could be invalidated if geological feedback mechanisms stabilize a water/land ratio close to unity. In a series of papers and contemporaneous with similar efforts \citep{cowan_water_2014, schaefer_persistence_2015, komacek_effect_2016}, H\"{o}ning and co-workers modelled water/land ratio outcomes considering plate tectonics-related feedback mechanisms in the Earth continental crust and water cycles \citep{honing_biotic_2014, honing_continental_2016, honing_land_2023}. These authors concluded that the cycle would tend to either maximize the continental (land) surface area covering almost the entire planet, or maximize the area of the ocean floor (thereby minimizing the land surface), but that a balanced distribution --- as on Earth --- is a comparatively rare outcome (see Section \ref{sec:deep-water-coupling}). 
The present Earth state would be marginally stable, perhaps kinetically stabilized by the decreasing mantle temperature, but would maximize continental crust-related geodynamic activity. Meanwhile, in the absence of any mechanisms regulating ocean volume \citep[see][]{karato_deep_2020}, there is no reason to expect that sea level remain below peak topography \citep{guimond_blue_2022}.

In summary, this picture of continental evolution raises the intriguing question of which if any components are inseparable from, and unique to, Earth's geodynamic history. 
Plate tectonics certainly set the thickness of Earth's continents, and may have stabilised their surface area \citep[e.g.,][]{Lenardic:2005, honing_continental_2016, cawood_continental_2019, cawood_secular_2022}. However, the relevant condition to forming TTGs in the Archean may be described as bringing hydrated crust to depth, and this effect could potentially be achieved through other mechanisms, such as delamination \citep[e.g.,][]{sizova_generation_2015}. Further, there exists evidence for land having emerged above sea level well before the accepted minimum age of subduction initiation \citep{chowdhury_continental_2024}. The striking implication of these ideas is that at least relatively modest volumes of felsic, buoyant landmasses on a planet may not necessarily require plate tectonics to form, although they may need water. 

In need of more clues, we return to the reputed 'continental analogues' of our nearest planetary neighbour. It is worth pointing out that the enigmatic tessera terrain-covered highlands on Venus, which morphologically, geodynamically, and perhaps even compositionally bear a strong resemblance to Earth's cratons \citep{hashimoto_felsic_2008, gilmore_virtis_2015} cover about 7\% of the surface of Venus \citep{Ivanov:2011}. Whether Venus' tessera-covered highlands truly are counterparts to the continental cores on Earth remains an open question, and may depend on any deep water cycle. Although at first glance, the tesserae and the cratons have similar areal extents today---the felsic continental interiors on Earth, the cratons, now occupy about 8\% of Earth's surface \citep{hasterok_new_2022}---whether there should be any significance to this fact is unclear, given that the margins of cratons are either structural (i.e., faulted) or covered by younger basins, implying that cratons had a greater extent in the past \citep[e.g.,][]{begg_lithospheric_2009}. Further, Venus' hypsometry is unimodal, without Earth's dichotomous elevations of oceanic and continental crust. Even so, the fact that there are highlands on Venus at all when the majority of the surface lies within only a few hundred metres of the mean planetary radius \citep{pettengill_pioneer_1980} means that, were there ever liquid water oceans on the planet (of a similar volume to Earth's), and assuming those highlands were present in at least comparable surface area to today, they would constitute `dry land' in a manner analogous to the subaerial continents on Earth.

\subsubsection{Other forms of topography}

There are additional mechanisms that can shape a planet's topography and thus, depending on the available liquid water volume at the surface, govern the water/land ratio. For example, large-scale tectonic processes, which principally manifest on Earth as vast lateral displacements, can have a vertical component of deformation, too. This process --- and especially the maximum achievable elevation --- is strongly controlled by the rheological properties of the lithosphere. Only cold, mechanically strong lithospheres can support high mountains or large volcanic edifices isostatically; in contrast, hot lithospheres deform and flow under such loads \citep{rey2008}. This rule is exemplified by the Voyager-era debate as to how such high mountains could exist on Io given observations of its high surface heat flow (implying a thin lithosphere under conduction). The seeming contradiction led \citet{oreilly_magma_1981} to suggest that Io's internal heat is transported by magma advection, thus giving rise to the heat-pipe theory.

Major thrust faults, formed through horizontal shortening and thickening of the crust, can readily generate topography several hundred to a few thousand meters in relief. On Earth, such thrust-fault systems are usually the result of tectonic plate convergence, forming complex thrust fault duplexes that are eroded into the familiar orogenic belts we recognize as the Alps, the Himalaya, and so on. Yet on Mercury, major thrust-fault systems standing $\sim$2--3\,km high and extending hundreds or even thousands of kilometres in length cross the planet \citep{byrne_mercurys_2014}, the result of secular cooling of the planet's interior that causes the solid planet to contract \citep{solomon_relationship_1977, tosi_thermochemical_2013}. Absent plate tectonics, large-scale crustal shortening remains possible---not to the extent on a stagnant lid world as on Earth, certainly, but enough to generate positive topography (and, by implication, spatial topographic variations in which water could pond as lakes or seas).

Whereas tectonic shortening of the crust produces positive relief, tectonic extension can similarly produce dramatic changes to the landscape, in the form of negative-relief topography. On Earth, the formation of (often vast) rift zones by tectonic plate divergence characterises seafloor spreading centres, but also occurs in continental settings; for example, within eastern Africa (cf. Fig. \ref{fig:dt}) and the southwestern United States. Some of these rift zones can have substantial negative relief (exceeding 1 km in the Basin and Range, USA), and considerable breadth (several tens of km in the case of the East African Rift). The driving mechanism for major rifting on Earth is plate tectonics, but crustal rifting is present on Venus and Mars, too \citep{anderson_primary_2001, Ivanov:2011, byrne_venus_2020}. On these worlds, the driving apart of the crust is likely the result of mantle upwellings and/or intrusive volcanic complexes under the stagnant lid tectonic mode --- with the Tharsis Rise on Mars, for example, having been explained by its underlying mantle superplume \citep[or more recently, transient plumes;][]{cheng_combined_2024}. Here, extension is not balanced by subduction but possibly by crustal thickening elsewhere, and so the total extensional strains are (much) lower than on Earth. Even so, topographic changes resulting from crustal rifting can be considerable.

Another non-tectonic mechanism, and one that operates widely on Earth albeit at relatively small spatial scales, is the formation of negative topography by karstic processes. Broadly speaking, karstic landscapes result from the chemical dissolution of soluble materials (chiefly limestone on Earth) by water, which takes advantage of existing fractures and other weaknesses in the crust. Karst occurs in a wide range of settings on Earth, and is characterised by the formation of often striking topography that includes towers, ridges, caves, and sinkholes --- the latter of which are frequently filled with water --- symbolised notably in the landscape around Guilin, China, for example. Karstic terrains have also been identified on Saturn's giant moon Titan, which in the present hosts voluminous deposits of liquid hydrocarbons that have presumably acted to erode the icy surface and subsurface in a manner akin to acidic water in carbonate settings on Earth \citep[e.g.,][]{malaska_labyrinth_2020}. Whether karstic process can operate on silicate exoplanets depends in great part on the composition (and solubility) of the crust, although carbonate rocks are not strictly necessary \citep{ford_karst_2013}, and of course on the availability of liquid water at the surface. Nonetheless, as a means to generate low-lying topography into which water can pond, karstic processes bear inclusion here.

The karstic processes described above, insofar as they create `negative topography', might be grouped into the broader process of erosion. Although erosion's net effect, over 100-Myr timescales, is to grind down mountains, on shorter time scales at which the elastic bending of the lithosphere and isostatic adjustment are relevant, erosion has complex effects on subcontinent-scale topography \citep[e.g.,][]{wolf_links_2022}. Physically removing mass --- erosional unloading --- through the action of glaciers, to exemplify the most important erosive agent on Earth \citep{dowdeswell_rates_2010} --- can cause uplift of the lithosphere locally as it rebounds \citep[e.g.,][]{champagnac_erosiondriven_2009, liu_unloading_2024}. The continents of a planet with glacial-interglacial cycles would bear the scars of glacier motion. Though the characteristic geomorphological features (e.g., drumlins, kettle lakes) are relatively short or shallow, they would be widespread, again guaranteeing the ubiquity of `little ponds'.

The processes discussed here are internally driven and reflect some interplay between the thermal evolution of the body, the convective vigour of the interior, and the mechanical strength (and composition) of the brittle lithosphere. Yet arguably the most effective mechanism for shaping a planet's topography is not internal at all. Impact bombardment has the demonstrated ability to dramatically alter a planetary body's landscape at all scales by excavating and redistributing crustal (and mantle) materials to form topographic lows and highs, often adjacent to one another, and thus creating undulating topography in which water could pond as lakes, seas, or even oceans. For example, at the very largest scales, impact cratering is responsible for the lunar South Pole–Aitken basin, the largest impact feature on the Moon \citep[e.g.,][]{zuber_shape_1994}; the mighty Caloris basin on Mercury \citep{murchie_geology_2008}; Mars' huge Hellas, Argyre, and Isidis basins \citep[e.g.,][]{schultz_new_1990}; and even, perhaps, the Red Planet's vast northern lowlands \citep[e.g.,][]{wilhelms_martian_1984, andrews-hanna_borealis_2008a}. Indeed, exogenous causes for the famous martian hemispheric dichotomy largely rely on one or more impacts so energetic as to have excavated and/or melted a large fraction of the surface. More recent explanations argue that although the lowlands are not an impact basin, the highlands are the result of crystallising a giant impact-induced magma pond in the south \citep{reese_impact_2011, golabek_origin_2011, cheng_combined_2024}. Although the past presence of an expansive northern ocean on Mars continues to be debated \citep{baker_ancient_1991, malin_oceans_1999, perron_evidence_2007, schmidt_circumpolar_2022, wu_probable_2024, li_ancient_2025}, there is little question that \textit{were} liquid water to exist on the surface in sufficient volumes to constitute a sea or even ocean, that water would certainly pond in the northern lowlands.

Far more certain is the ponding of liquid water in Mars' smaller impact features, evinced by the lacustrine sedimentary deposits in Gale \citep{edgar_lacustrine_2020} and Jezero \citep{goudge_assessing_2015} craters, the field sites of the robotic Curiosity and Perseverance geologists, respectively. These craters are much smaller in size ($\sim$50–-150 km wide) than the giant basins listed above (which all exceed 1,500 km in diameter), and so could much more readily host standing bodies of water of even modest volume. Of note, impact features (on rocky worlds) can be isostatically compensated and so do not require dynamic support from below to form or be sustained, with even the very largest basins being supported by membrane stresses,\footnote{That is, stresses that are uniform with depth in the lithosphere, which behaves elastically.} and thus these features can persist over geological time. Viscous relaxation of the crater form on icy worlds is well documented \citep[e.g.,][]{parmentier_viscous_1981}, but much less so on silicate bodies.

There are clearly no shortage of processes that can generate variations in topography that deviate from some mean, thus allowing the ponding of water into lakes, seas, or even oceans---assuming such water volume is present. On the basis of what we see in the Solar System, it seems to be far more probable to produce and sustain high- and low-standing topography than it is to form and hold onto an ocean. Whether that is the case for extrasolar worlds is discussed below.

\subsection{Planetary surface water inventories}\label{sec:water-budget}
\subsubsection{Estimating planetary bulk water budgets}\label{sec:water-delivery}

In practice, inferences about an exoplanet's mass fraction of highly volatile materials (water and ices, e.g.) can come from its measured bulk density because volatiles are less dense than rock and iron (see Baumeister, Miozzi et al., 2026, this topical collection, for an extensive discussion). For those planets \textit{less} dense than rock and iron, a substantial fraction of light material is required to make up the density deficit. However, even if we could know that this light component were water in condensed form on the surface of the planet, the amount of water required to make an Earth-mass planet measurably less dense \citep[i.e., by 0.3 g\,cm$^{-3}$ by analogy to the TRAPPIST-1 system;][]{agol_refining_2021} is about 2--3\% of the planet mass, or 180--250 km deep, for example.\footnote{If this water were instead dissolved in a fully-molten mantle of a young, hot planet, an even higher water mass fraction would be required to affect the bulk density signature \citep{dorn_hidden_2021, luo_majority_2024}. A \textit{solid} mantle is unlikely to be able to store enough water to increase its radius beyond the observational uncertainty, given lower water concentrations at water saturation \citep{shah_internal_2021}.} This volume of water would necessarily result in an ocean-covered world, as it well exceeds the supportable topography \citep{guimond_blue_2022}. Even wielding infinitely-precise measurements, larger ocean masses could be masked by a larger bulk iron content, absent degeneracy-breaking prior knowledge (such as a hot equilibrium temperature\footnote{The equilibrium temperature of a planet is the temperature at which it radiates away all of the stellar energy it receives, less a nominal fraction reflected. The presence of greenhouse gases would raise the actual surface temperature above this equilibrium temperature.} precluding surface oceans or ice). Therefore, although bulk density observations may identify distinctly volatile-rich planets as such (Baumeister, Miozzi et al., 2026, this topical collection), we do not expect to be able to use this technique to infer the ocean mass fraction of a potentially `blue marble' planet. 

The plausible water budgets of cool, potentially-rocky planets could be informed instead by theory encompassing planet formation, metal-silicate differentiation (i.e., segregation of H and O into a metal core), and irreversible loss of the atmosphere to space (on the latter, see Kubyshkina et al., 2026, this topical collection). To first order, whether the main building blocks of a planet originate outside or inside the snow line in a protoplanetary disk (corresponding to the temperature below which water condenses) controls whether this planet likely forms with large ($\gg 1$ wt.\%) or small ($\ll 1$ wt.\%) water mass fractions \citep[e.g.,][]{tian_water_2015, lichtenberg_bifurcation_2021}. 
In detail, however, the complexities and stochastic nature of the aforementioned processes --- as well as others not considered here, such as orbital migration --- mean that predicting the precise water inventory of an exoplanet inside the snow line is likely to be impossible.

Nonetheless, there are a number of potential, non-mutually-exclusive sources of water to the growing Earth and other planets inside their disk's snow line:
\begin{enumerate}
    \item \textit{Wet accretion:} The primitive building blocks of planets may themselves contain some fraction of hydrogen by mass. In the inner regions of the Solar System, these building blocks may be represented by enstatite chondrites \citep{javoy_integral_1995}. Although previously thought to be dry, recent measurements by 
    \citet{barrett_source_2025} have shown that enstatite chondrites do contain H bound to S, enough to supply up to 10 ocean masses\footnote{An ocean mass amounts to about 1.34 $\times 10^{21}$ kg, equivalent to 333 wt. ppm of the bulk silicate Earth.} from 1 Earth mass of building blocks. In addition to the enstatite chondrite contribution, carbonaceous chondrite material from the wetter, outer Solar System could make up $\sim$2--5\% of Earth's mass, corresponding to several oceans \citep[based on Xe isotopes;][]{cassata_refined_2025}. A potential issue with accreting planetary water from undifferentiated chondrites is that these building blocks would have not experienced much heating: else they would have differentiated, degassing water in the process. More strongly-heated protoplanets would make drier planets \citep{lichtenberg_water_2019, sanderson_differences_2024}.
    \item \textit{Late(r) accretion:} After the bulk of the planet's mass has accreted and differentiated, water is delivered to its surface via impacting material scattered in from beyond the snow line \citep[e.g.,][]{morbidelli_source_2000, hartogh_oceanlike_2011, alexander_provenances_2012, marty_origins_2012, mandt_nearly_2024}. Isotope constraints point to at least some contribution to the bulk Earth from outer Solar System material \citep[e.g.,][although the exact contributions remain unresolved]{burkhardt_terrestrial_2021}. The important distinction between this process and the one above is that the cumulative water delivered through smaller impacts depends on highly-stochastic dynamical scattering, sensitive to, for example, the migration history of the giant planets \citep{raymond_making_2004}. Meanwhile, accreting water at this post-differentiation stage guarantees that such delivered water is not permanently sealed in an iron-alloy core.
    \item \textit{Nebular ingassing:} Hydrogen gas of stellar nebular origin,\footnote{Whereas there is no direct evidence of an enduring primordial hydrogen atmosphere on Earth, it has been proposed that many observed rocky exoplanets today are the remnants of initially more gas-rich planets that have lost their hydrogen atmospheres \citep{owen_hydrogen_2020}. However, since the photoevaporative process that removes hydrogen would work less efficiently on cooler planets, it remains to be seen whether more-temperate rocky exoplanets have experienced something similar.} upon reacting with a primordial magma ocean, could be oxidised to produce water if it lingers long enough \citep{ikoma_constraints_2006, sharp_hydrogenbased_2013, sharp_nebular_2017, olson_hydrogen_2018, olson_nebular_2019, kimura_formation_2020, kite_water_2021,kimura_predicted_2022, young_earth_2023, krissansen-totton_erosion_2024}. For example, \citet{olson_nebular_2019} estimated that, in theory, Earth could have acquired and retained at least an oceans' worth of water through nebular ingassing alone. This estimate would be lower for a less-oxidising magma ocean; in this way such nebular water production is also tied to the relative timing of core formation.
    \item \textit{Direct accretion of H$_2$O vapour:} Water-bearing material drifting inward across the snow line could sublimate, whereupon the water vapour viscously diffuses under the influence of a growing planet's gravity \citep{kral_impactfree_2024, houge_smuggling_2025}.
\end{enumerate}

\paragraph{The bulk water budget at birth}

In sum, the multitude of processes that might deliver and/or make available water to inner regions of protoplanetary disks suggests that it need not be difficult for some rocky planets to gain rather substantial initial water inventories. Meanwhile, planetary differentiation ensues; possibly-large fractions of hydrogen partition into the metal that initially separates from silicate material \citep[hydrogen favours metallic melt over silicate melt at terrestrial planet core-formation pressures; e.g.,][]{okuchi_hydrogen_1997, li_earths_2020, tagawa_experimental_2021, luo_majority_2024, liu_hydrogen_2024}. This water equivalent probably becomes locked in the core forever.\footnote{The core is potentially the largest H reservoir in the interior of the Earth; 51 oceans-worth of hydrogen have recently been proposed by \cite{liu_hydrogen_2024}. Water `locking' in the core is helped by the small diffusivity of H in lower mantle rock.} As for the remaining water, we now believe water to be highly soluble in primordial magma oceans, even miscible above a few GPa, such that (temporary) sequestration of this water is quite feasible \citep[e.g.,][]{bureau_complete_1999, mibe_second_2007, dorn_hidden_2021, bower_retention_2022, sossi_solubility_2023}. 

It then becomes a different question of whether the bulk silicate planet's water can be retained through the early evolution of the planet. Answers to the water retention question are tied to the duration of a primordial magma ocean and its trade-off with the timing of escape of hydrogen from the upper atmosphere \citep[e.g.,][]{zahnle_evolution_1988, lebrun2013, hamano_emergence_2013, miyazaki_inefficient_2022}, subject further to unpredictable catastrophic events such as giant impacts \citep[e.g.,][]{biersteker_losing_2020a,  saurety_impactinduced_2025}. On the one hand, a long-enduring magma ocean, continuously degassing volatiles,\footnote{In magma ocean models, the degassing rate of water from the magma surface is typically controlled by a vapour pressure equilibrium (Henry's Law). If atmospheric escape drives the partial pressure of water to continuously decrease, more water will be continuously degassed to maintain equilibrium, barring some other bottleneck to degassing \citep[such as in][]{salvador_convective_2023, walbecq_effect_2025}.} would be exposing more water vapour to extreme UV radiation from the host star \citep{hamano_emergence_2013, kite_exoplanet_2020, salvador_magma_2023}. On the other hand, the very high potential of magma to hold water means that, \textit{for as long as this magma ocean has not mostly solidified}, it could easily be protecting many ocean-masses of water from such radiation \citep{dorn_hidden_2021, bower_retention_2022, sossi_solubility_2023, moore_role_2023, nicholls_magma_2024, maurice_volatile_2024}. As the magma ocean cools and crystallises, the capacity of the mantle reservoir markedly decreases because water is much less soluble in silicate \textit{crystals} compared with melt (see Section \ref{sec:nam-capacity}). The inevitable decrease in reservoir capacity implies that a large fraction of water must go somewhere else, presumably aboveground \citep[e.g.,][]{tikoo_fate_2017, krissansen-totton_predictions_2022a, miyazaki_inefficient_2022, salvador_magma_2023}, by which time the star may or may not have become less active. In any case barring total dessication, the newly-crystallised mantle rock still traps some of the magma ocean's hydrogen according to its finite solubility in this rock, as could any interstitial melt \citep{tikoo_fate_2017, miyazaki_inefficient_2022, miyazaki_wet_2022, salvador_magma_2023}. Thus there are potential avenues for at least \textit{some} internal water retention even after the magma ocean phase effectively ends. 

Massive water losses do seem particularly hard to avoid for planets on close-in orbits around M-dwarfs --- i.e., most of the best-observed rocky exoplanets to date --- given inferred high rates of atmospheric escape even to the present (see Kubyshkina et al., 2026, this topical collection). Nevertheless, astronomers are seeking to empirically test the retention of volatiles on rocky planets by looking for signs of atmospheres (see Lustig-Yaeger et al., 2026, this topical collection). 

Do we expect a young planet's hydrosphere to condense as oceans? Water not containable in the mantle may be quickly degassed as steam. This steam condenses if the surface and atmosphere can sufficiently cool down, implying that greenhouse gases such as CO$_2$ must not be \textit{too} abundant \citep[e.g.,][]{salvador2017, salvador_magma_2023, miyazaki_inefficient_2022}, whereas star-planet distance (planetary insolation) may play a key role \citep{turbet_day_2021}. Conditions were evidently met on early Earth \citep[e.g.,][]{mojzsis_oxygenisotope_2001}, but the same is not obvious for all planets, as exemplified in the active debate around an early temperate phase on Venus \citep[e.g.,][]{way_venusian_2020, turbet_day_2021, westall_habitability_2023}.\footnote{Although this review does not cover how and when plate tectonics can initiate, we note that ocean condensation is probably important \citep[e.g.,][]{miyazaki_wet_2022}.} 


\subsubsection{The deep water cycle: geodynamics and mineralogy}\label{sec:deep-water-cycle}


\begin{figure}
    \centering
    \includegraphics[width=1\linewidth]{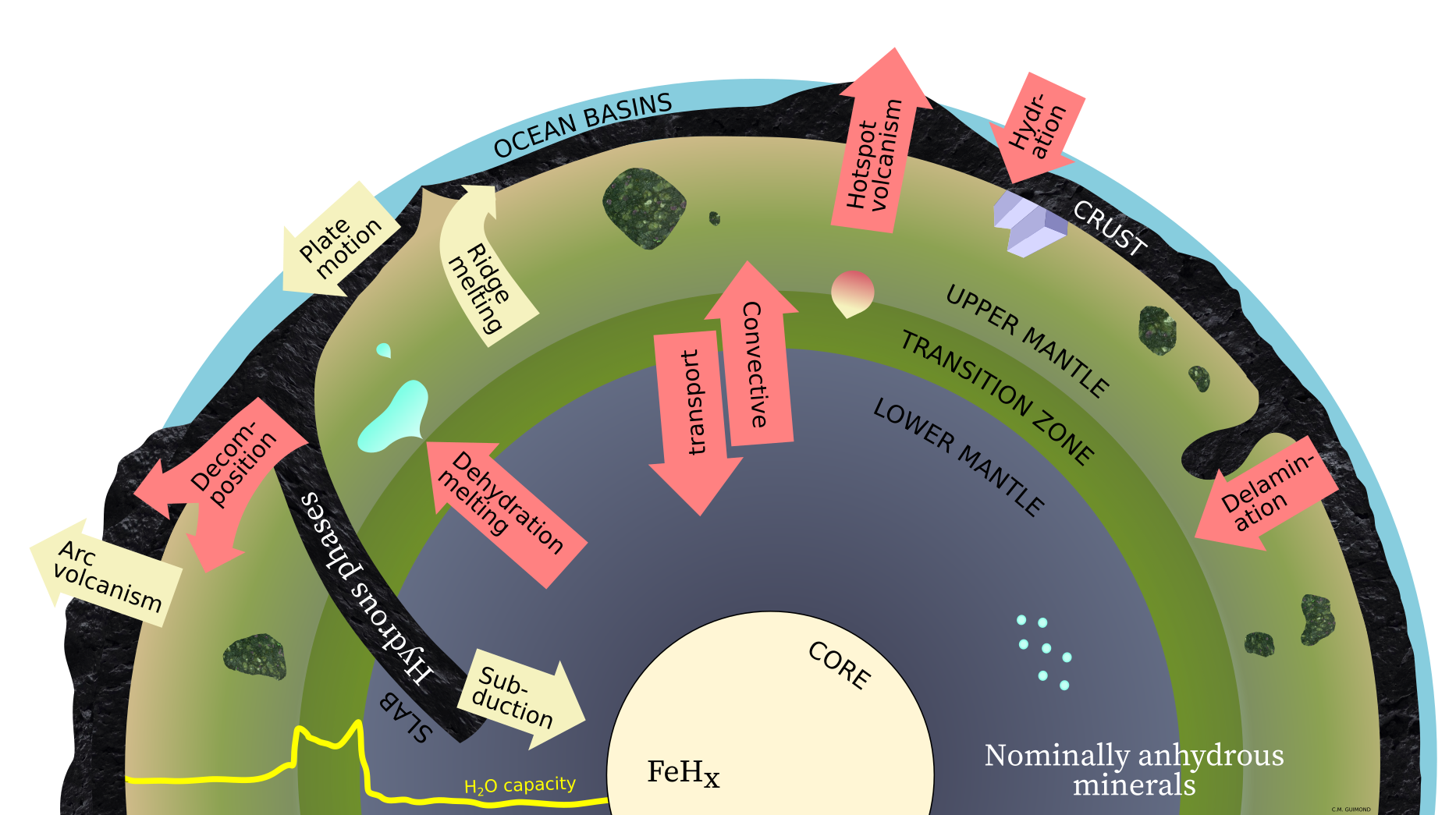}
    \caption{Elements of the deep water cycle on rocky planets that are discussed in this chapter. Reservoirs are labelled in sans serif font. The dominant form of `water' in certain reservoirs is noted in serif font. The major fluxes potentially possible on rocky planets in general are marked in red arrows, whereas the fluxes operating only under plate tectonics are marked in yellow arrows. The profile in the bottom left corner indicates the relative water concentration at water saturation in nominally anhydrous minerals for an Earth-like mantle composition. Note geometry is not to scale.}
    \label{fig:cycle}
\end{figure}

A planetary body's bulk water content is mostly set during its formation, but the distribution of water between the surface and the interior can vary between planets and in time \citep[e.g.,][]{williams_water_2007, karato_water_2015, salvador2017, miyazaki_inefficient_2022}. As the previous section suggested, water is feasibly present inside the body from its early stages. Given billions of years, this interior water reservoir is carried through the mantle as it convects, and could slowly empty to the surface through subsequent melting and volcanic outgassing. The action of plate tectonics changes the loci of outgassing, but volcanism in the first place does not require plate tectonics (Fig. \ref{fig:cycle}). Phenomena analogous to Earth's hotspot volcanoes (e.g., Hawai'i) are evidenced by Olympus Mons (and major plains volcanism on all single-plate worlds). 

Less certain is the ubiquity of water ingassing to the mantle. On modern Earth, the ingassing mechanism takes the form of subduction of crust containing hydrated minerals, so Earth's deep water cycle is completed \citep[e.g.,][compare Fig. \ref{fig:cycle}]{jacobsen_Earth_2006, rupke_geological_2024}. About $10^{12}$ kg of water enters Earth's subduction zones each year, largely as hydrous minerals and pore water in sediments and hydrated basaltic crust \citep[and references therein]{karato_water_2015}. Indeed, Earth is believed to have been in a state of net water ingassing to the mantle for a large part of its history \citep[e.g.,][and references therein]{ito_alteration_1983, jarrard_subduction_2003, vankeken_subduction_2011, karlsen_deep_2019, korenaga2021}. The geological record suggests an intriguingly constant freeboard (or, sea level) during the last $\sim$500\,Myr \citep[e.g.,][]{parai_large_2012}, perhaps hinting at a feedback mechanism at play. 

We can look to the Solar System for clues about water ingassing under other tectonic modes. There are no active subduction zones on Mars, but some ingassing mechanism is not ruled out completely. Petrologic modelling suggests that Mars' small size and iron-rich crust could allow that crust, upon hydration (consuming surface liquid water), to remain dense enough to remain buried under new volcanic lava flows, eventually sinking to the upper mantle, bringing hydrated minerals with it \citep{wade_divergent_2017}. 
Although little about Venus' water cycle is known, evidence for compositionally-diverse lava indicates some amount of heterogenous processing, perhaps mediated by volatiles. Hence \citet{ElkinsTanton:2007} demonstrated with a geodynamic model that gravitational instabilities could drag parts of a volatile-bearing lithosphere back to the upper mantle in this scenario. It should be noted that in reality, so-called tectonic modes (hence available ingassing mechanisms) need not be neat nor permanent \citep{lenardic_diversity_2018}. For example, water may be returned to the mantle during an episode of resurfacing for an otherwise fairly immobile lithosphere \citep[e.g.,][]{vezinet_growth_2025}.

Ingassing is important because in its absence, and given some degree of outgassing, a temperate planet's ocean volume would always increase with time, barring rapid atmospheric escape (or impact blowoff) or a climate catastrophe. Some planets may also have limited-to-no \textit{out}gassing --- hence no flux of water at all --- if wet buoyant magma does not form under a very thick lithosphere, as predicted for more massive rocky planets without plate tectonics \citep{kite_geodynamics_2009, Dorn:2018, ortenzi2020}. 

\paragraph{Water capacities of minerals}\label{sec:nam-capacity}

Deep water cycling and partitioning must obey physical limits to the capacities of various water reservoirs in the planet. Although the surface water reservoir is essentially limitless,\footnote{However, any steam atmosphere could not extend beyond a planet's gravitational sphere of influence.} water has natural solubility limits in silicate minerals and melts. Mantle and crust minerals can store water in substantial quantities \citep[e.g.,][]{bolfan-casanova_water_2005, ohtani_hydration_2021, mosenfelder_hydrogen_2024} either as stoichiometrically hydrous minerals (e.g. serpentine, amphibole), or as non-stoichiometric -H or -OH in nominally anhydrous minerals or NAMs (e.g., olivine, orthopyroxene), the latter of which make up the bulk of silicate mantles. The `water-equivalent' storage capacity in NAMs is related to the nature of the crystal lattice substitution enacted by -H or -OH, and depends on pressure, temperature, and redox conditions. Estimates of water content in the bulk present Earth vary widely, from one to an extreme 100 ocean masses \citep{peslier_water_2017}, but a recent estimate arrives at 2.6 to 8.3 ocean masses stored in the mantle and crust \citep[generally below water saturation;][]{ohtani_hydration_2021}. Of this total, only 0.3 to 0.4 ocean masses are in the crust ($\sim$0.21 in the continental crust), largely as stoichiometric hydrous minerals rather than NAMs, at concentrations of around 15,000--20,000 wt. ppm \citep{bodnar_whole_2013, peslier_water_2017}. 

In part because different NAMs are stable at different mantle depths according to temperature and pressure, and also because water concentrations at water saturation in NAMs are temperature-dependent, `water' is not expected to be distributed homogeneously inside a planet. The typical assemblage of upper-mantle minerals will store about 10$^2$ to 10$^3$ wt. ppm water equivalent at water saturation \citep[as compiled and re-fit by][]{dong_constraining_2021}. 
Meanwhile, in the 250-km-thick transition zone between the upper and lower mantles, the stable NAMs wadsleyite and ringwoodite can accommodate hydrogen equivalent to 1.8--2.3 $\times 10^4$wt. ppm and 1--1.25 $\times 10^4$ wt. ppm water, respectively 
\citep{inoue_water_2010, tschauner_ice_2018}. Wadsleyite/ringwoodite layers are predicted to be thermodynamically stable across many rocky exoplanets, implying that transition zone-like structures are common, with potential to hold much water, if filled --- dominating the water storage capacities of planets up to $\sim$1.5 Earth masses \citep{guimond_mantle_2023}. The water storage capacity of the lower mantle is debated, although it is accepted to be much smaller than that of the transition zone, at perhaps 100--2000 wt. ppm \citep[e.g.,][]{karato_water_2015}. \citet{shah_internal_2021} and \citet{guimond_mantle_2023} estimated the water storage capability of rocky planets depending on mass and mineralogy, considering reasonable temperature and pressure variations with depth. \citet{guimond_mantle_2023} found that mineralogically-average rocky planet mantles would saturate at between about 2 and 8 Earth ocean masses, for planet masses from 0.5 to 5 Earth masses respectively, whereas \citet{shah_internal_2021} find higher values since they used a much higher water capacity of postperovskite (3 wt.\%), the high-pressure silicate dominating deep super-Earth mantles. Again, of course, mantles need not be saturated in water. Geophysical evidence reviewed in \cite{ohtani_hydration_2021} suggests that Earth's transition zone and upper mantle are quite far from water saturation (100--200 wt. ppm water in the upper mantle), and further, that the actual distribution of water in the transition zone itself is not homogeneous. 

How do the water capacities of minerals affect how they move water into and through a planet's deep interior? Hydrous minerals --- chlorite, amphibole, serpentine --- easily carry up to tens of wt.\% water, but they tend to only be stable in cool regions of the interior, such as in the crust, and on a plate tectonics world, in subducted slabs of oceanic lithosphere. High-pressure hydrous phases \citep[the ABC phases of][]{ringwood_high-pressure_1967} can exist at deep lower mantle pressures of up to 60\,GPa, on cool slab temperature-pressure profiles (not on typical mantle adiabats). These slabs where present --- and mantle material viscously coupled to the slabs --- can be major reservoirs of water, albeit transitory at geological time scales \citep[e.g.,][]{goes_subduction_2017, ohtani_hydration_2021, keppler_subduction_2024}. In this way, subducted hydrous minerals play a key role in Earth's water ingassing mechanism. 
Possibly large yet difficult-to-estimate fractions of this subducted water are expelled at shallow depths, however \citep[e.g.,][]{honing_biotic_2014}. The remaining water, partly stored in serpentines and other hydrous minerals, then reaches the source depths of continental crust rock, where more hydrous minerals break down still, releasing water that is consumed to make new continental crust. In the end, a relatively small fraction of subducted water reaches the mantle itself. Some of the water will be released to the mantle wedge above the slab, where it may be incorporated into local NAMs; the descending slab may drag part of the hydrated mantle to even greater depth where water is eventually incorporated into lower mantle NAMs \citep[e.g., perovskites, periclase, stishovite;][Fig. \ref{fig:cycle}]{novella_deep_2024}. 
As a consequence, whereas the residence time of water in the very deep mantle may exceed 1 billion years, shallow dehydration operating on much shorter timescales could ultimately govern the upper-mantle water budget \citep{nakagawa_numerical_2023}.

Differences in the water storage capacity between the mantle layers and phase transitions at the top and the base of the transition zone may have important consequences for the dynamics of Earth's mantle, such as initiating local melting where upwelling rock becomes supersaturated and dehydrates upon a mineral phase change \citep[e.g.,][figure \ref{fig:cycle}]{bercovici_whole_2003, karato_water_2015, ohtani_hydration_2021, rupke_geological_2024}. The difficulty of redistributing water in the mantle through convection has been discussed by \cite{richard_transition_2002} and \cite{karato_water_2015}. For example, the endothermic phase transition from ringwoodite to bridgmanite and magnesiowüstite can cause a subducting slab to stall \citep[e.g.,][]{goes_subduction_2017, keppler_subduction_2024} and inject water into the transition zone, rather than transporting it to the lower mantle. 

\paragraph{Coupling between mantle water, dynamics, and continental crust production}
\label{sec:deep-water-coupling}

\begin{figure}
    \centering
    \includegraphics[width=0.7\linewidth]{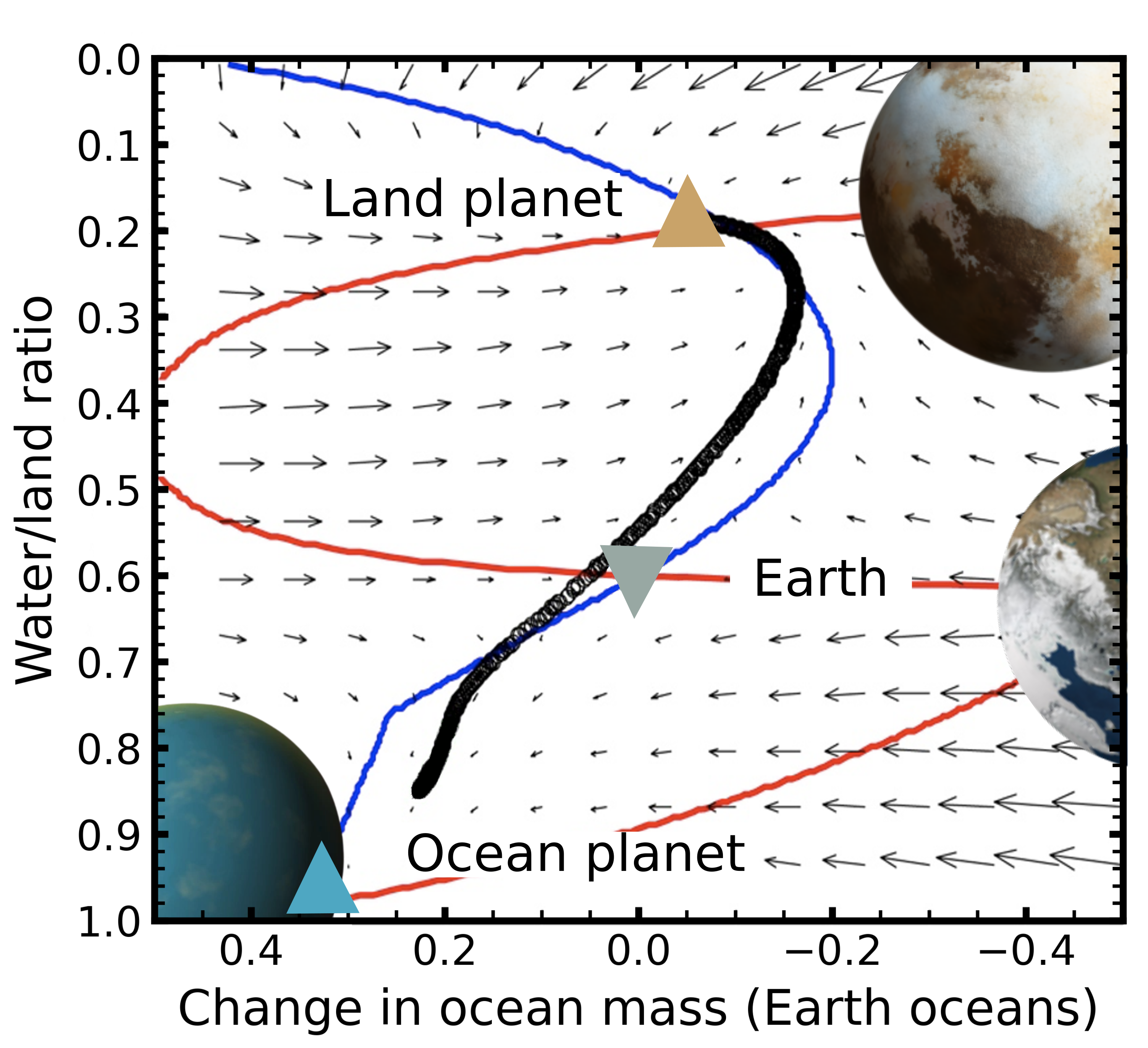}
    \caption{Phase plane spanned by the surface ocean mass (expressed as the difference from one Earth ocean mass) and the coverage of the Earth's surface by continental crust, calculated by \citet{honing_continental_2016} for present-day mantle temperatures. 
    Local arrows are shown pointing in the direction of time evolution. Also shown are lines along which the rates of change of continental coverage (red) and ocean mass (blue) are zero. These lines intersect at three points (triangle markers) where both rates of change are zero, indicating equilibrium states. At the top fixed point (brown triangle; `land planet'), the planet is at a stable equilibrium, mostly covered by continents; net $\sim$0.1 ocean masses have been ingassed to the mantle. At the lower fixed point (blue triangle; `ocean planet'), the planet is at another stable equilibrium, almost completely covered by oceans; net $\sim$0.4 surface ocean masses have been gained from mantle degassing. The middle fixed point (grey triangle; `Earth') is a saddle point, stable with respect to ocean mass but unstable with respect to continental coverage. The black circular markers indicate the endpoints of the 4.5 Ga evolution calculations, which do not necessarily reach equilibrium states. About 80\% of the endpoints cluster around the `land planet' fixed point; the lower fixed point is never reached because the mantle becomes too dry, cold, and viscous. Only a few percent of models end close to the `Earth' saddle point. The models are initialised with randomly chosen temperatures between 1800 and 2100 K and with an inventory of one ocean mass in the mantle.}
    \label{fig:phase plane}
\end{figure}

Water can have a major effect on mantle flow by reducing the activation enthalpy for creep \citep[e.g.,][]{griggs_quartz_1965,hirth_rheology_2003, karato_pressure_2003, karato_water_2011}. Further, the thermal conductivities of olivine \citep{chang_hydration_2017}, ringwoodite \citep{marzotto_effect_2020}, and lower mantle mineral assemblages \citep{hsieh_spin_2020} have also been reported to decrease with increasing water content, and NAMs' water storage capacities themselves exhibit a strong temperature dependence \citep{karato_water_2015}. The dependence of mantle flow on water content sets up potential feedback mechanisms \citep[e.g.,][]{mcgovern_thermal_1989, sandu_effects_2011, honing_biotic_2014, honing_land_2023}. An increase in mantle water content will lower the viscosity 
so the mantle flows more vigorously. This enhanced flow will, in turn, increase the subduction rate (i.e. the ingassing rate) in a plate tectonics regime and amplify the initial increase in mantle water content. However, the faster mantle flow will also promote more water outgassing, dampening the positive feedback loop. This water-cycling process is coupled to the continental crust cycle through the contribution of wet, continent-derived sediments to water ingassing at subduction zones, and through the key role of water in the generation of granitic continental crust \citep{honing_land_2023}. The growth in land volume in these models compensates for the loss of surface ocean volume and tends to keep the average ocean depth constant. It is probably not surprising that these complex and intersecting feedback loops could produce bifurcations in both ocean volume and ocean basin capacity (topography); there may be more than one equilibrium solution for the same planetary conditions (compare Fig. \ref{fig:phase plane}). 

\paragraph{What controls water outgassing rates?}

\setlength{\tabcolsep}{4pt} 
\begin{table}[h]
\centering
\begin{tabular}{lccccc}\toprule
Reference & Model & Mantle water & $f_{{\rm O}_2}$ & Mode & Flux\\
          &       & (wt. ppm) & ($\Delta$Earth) & & (OM\,Gyr$^{-1}$) \\\midrule
\citet{sandu_effects_2011} & PCT & [689, 2068] & 0 & PT & [0.01, 8] \\
\citet{komacek_effect_2016} & PCT & [470, 720] & 0 & PT & [0.01, 5] \\
\citet{krissansen-totton_predictions_2022a} & PCT & [30, 30000] & [$-$4, 3] & PT & [$\sim$0, 16] \\
\citet{honing_land_2023} & PCT & $\sim$80 (UM) & 0 & PT & [0.03,0.07] \\
\citet{Tosi:2017} & PCT & [250, 1000] & 0 & SL & [0.003, 0.03] \\
\citet{miyazaki_inefficient_2022} & PCT & [100, 1000] & 0 & SL & [0.007, 0.2] \\
\citet{spaargaren_influence_2020} & MLT & 695 & 0 & SL & 0.7 \\
\citet{spaargaren_influence_2020} & MLT & 695 & 0 & PT & \textless 4 \\
\citet{guimond_low_2021} & 2D & [50, 450] & [$-$6, 0] & SL & [0.004, 0.3] \\
\citet{nakagawa_numerical_2023} & 2D & \textless 10000 & 0 & PT & [0.008, 8] 
\\\bottomrule
\end{tabular}
\caption{Selected volcanic H$_2$O outgassing fluxes for a $\sim$1 Earth mass planet from mantle thermal history models. Variations in tabulated outgassing fluxes are not only due to differences across the columns listed here; see the original publications for details. Note earlier publications of a similar model are omitted (as are those which do not clearly provide an H$_2$O outgassing flux). The numbers in the square brackets, separated by commas, represent ranges of the assumed parameters and results. $f_{{\rm O}_2}$ is the mantle oxygen fugacity expressed as the log difference from modern Earth. OM: Ocean Masses ($1\,{\rm OM} = 1.34 \times 10^{21}\,{\rm kg}$); PCT: 0/1D parameterised convection theory; MLT: 1D mixing length theory; PT: plate tectonics; SL: stagnant lid; UM: upper mantle.}
\label{tab:outgassing}
\end{table}

Rates of volcanic outgassing are an outcome of complex feedback loops involving both the surface and interior of the planet. Nonetheless, we can identify certain factors or conditions setting how efficiently a mantle outgasses water. The tectonic mode of the planet certainly appears crucial. Observing rates of volcanic melt production has not been possible for other rocky bodies outside the plate tectonics mode, excepting the heat-pipe archetype Io \citep[where rates of crust production can be inferred from heat flux observations;][]{oreilly_magma_1981}, so our estimates of volcanism rates in other tectonic modes come largely from inferences made on the basis of photogeological mapping and geodynamic modelling. 

Across different models, stagnant-lid scenarios result in rates of volcanism diminished by approximately an order of magnitude compared to plate tectonics scenarios \citep{kite_geodynamics_2009, oneill_mantle_2013, noack_waterrich_2016, Noack:2017}. Part of the difference in magma production rates can be explained by the ranges of depths where upwelling mantle will melt: under a very thick ($\sim$100\,km), extremely viscous (`stagnant') global lid, upwelling stops at relatively high pressures, limiting adiabatic melting. Conversely, mantle convection essentially reaches the surface (at mid-ocean ridges) in the plate-tectonics mode. However, this explanation does not always extend to outgassing of water. Because water solubility in silicate melts increases quickly with pressure, water can remain largely dissolved in these melts at Earth's mid-ocean ridge pressures. Water outgassing from Earth's mantle is dominated instead by volcanic arcs, where melting is induced by adding water (Fig. \ref{fig:cycle}).

Indeed, the pressure-dependence of water solubility may be an important accident of nature that controls water outgassing. In an influential study, \citet{gaillard_theoretical_2014} suggest that Venus' $\sim$100-bar atmosphere may prohibit the outgassing of any mantle water. Hence, through its modulation of the ocean depths overlying possible volcanic vents, the water/land ratio itself would also regulate water degassing. Water exsolution can be affected at ocean depths similar to terrestrial mid-ocean ridges, and has been predicted to virtually shut off for oceans deeper than 10\,km \citep{krissansen2021}.

As for overall rates of volcanism --- the surficial expression of melt --- some of the difference between tectonic modes is due to differences in the stress state of the lithosphere, among other factors. 
This discussion has not covered all of the factors that can affect water outgassing rates; an incomplete list follows. A 2D convection model of Archean Earth in the stagnant-lid mode \citep{guimond_low_2021} found water outgassing rates correlating most strongly with the initial bulk mantle water content, the fraction of melt that extrudes as opposed to stalling deep in the crust, and mantle oxygen fugacity (increasing the gas phase H$_2$O/H$_2$). Nonetheless, for H$_2$ outgassing to be appreciable, melt oxygen fugacities must be lower by at least 4--6 orders of magnitude compared to modern Earth \citep{brachmann_distinct_2025}. Planet mass might be expected to influence outgassing at many stages of the process, through direct effects (higher gravity, interior temperatures, convective vigour) and indirect ones (likely tectonic modes). In 2D models, stagnant lid outgassing rates have been shown to increase with planet mass to a maximum at $\sim$3--4 Earth masses, beyond which steep interior pressure gradients effectively shut off melting \citep{Noack:2017, ortenzi2020}; an appropriate treatment of melting in 1D can reproduce this behaviour \citep{kite_geodynamics_2009}. Melting the mantle requires heat, so planets with stronger internal heating --- e.g., endowed with a greater concentration of radiogenic heat producing elements such as U, Th, and K compared with Earth --- would experience more volcanism and outgassing, for longer \citep{frank_radiogenic_2014, nimmo_radiogenic_2020, Unterborn:2022}, in a generally linear correlation best seen when low reference viscosities reduce the lag between internal heating and convective heat transport \citep{moore_efficiency_2015, Unterborn:2022}. The greatest rates of volcanism in the solar system are observed on Io, which is tidally heated to the extent that its surface heat flow is $\sim$20$\times$ Earth's despite its small size \citep[e.g.,][]{oreilly_magma_1981}. Any convection-outgassing model will be sensitive to the assumed mantle solidus and the pressure at which melt becomes neutrally buoyant \citep{stolper_melt_1981}, pointing to the important role of experiments in constraining these values in particular for non-Earth-like mantle compositions \citep{kiefer_effects_2015, brugman_experimental_2021}.

In summary, for a water-bearing mantle to outgas water efficiently, two main characteristics are favoured: \textit{(i)} a relatively thin lithosphere to promote adiabatic melting; and \textit{(ii)} a relatively thin atmosphere and/or ocean to promote water exsolution from magma. The fact that so many other possibly relevant conditions are unconstrained means that quantitative predictions of water outgassing rates would be probabilistic. To expose the range of such predictions and give context to the rest of this article, Table \ref{tab:outgassing} compares the water outgassing rates from a selection of mantle convection models. Interestingly, stagnant-lid convection models produce a similar maximum water outgassing rate of 0.2--0.3 ocean masses per Gyr across different model setups and assumptions. Plate tectonics models consistently produce a larger maximum water outgassing rate of 1--10 ocean masses per Gyr. Plate tectonics models tend to use more diverse parameterisations; notably, where outgassing depends solely on mantle temperature in \citet{schaefer_persistence_2015} and \citet{komacek_effect_2016}, outgassing rates drop to order $\sim$0.01--0.1 ocean masses per Gyr. We offer the provisional conclusion that long-term water outgassing rates far exceeding 10 ocean masses per Gyr would require unusual conditions that are not normally considered in rocky planet thermal history models.

\section{Consequences of water versus land surfaces on climate and life}\label{sec:consequences}
   

The whole spectrum of water/land ratios is possible on rocky worlds: by now we hope to have moved beyond viewing modern Earth as default. This notion matters because planets' land and water masses perform several important functions in controlling and regulating both surface temperature (and hence the planet's location with respect to the classic circumstellar habitable zone) and biological productivity. The importance of land versus ocean surfaces is also, notably forefronted in origins of life scenarios \citep[e.g.,][]{deamer_urability_2022}. For example, Darwin's hypothesis that life begins in `warm little ponds' --- such as hot springs on subaerial volcanic platforms --- is finding increasing support in geological evidence \citep[e.g.,][]{djokic_earliest_2017, vankranendonk_origin_2021}, prebiotic chemical modelling \citep[e.g.,][]{pearce_origin_2017}, and laboratory experiments \citep[e.g.,][]{Mulkidjanian:2012, deguzman_generation_2014}. The major competing idea, origin of life at hydrothermal vents on the seafloor, meanwhile requires the existence of oceans and a geologically active seafloor \citep{corliss_hypothesis_1981}. Whether early Earth's actual subaerial land fraction was minute \citep{chowdhury_continental_2024} or substantial \citep{armstrong_radiogenic_1981, rosas_rapid_2018} then becomes highly relevant to either scenario's favourability. Heated debate continues as to the environment in which life began on this planet: warm little ponds, hydrothermal vents, or somewhere else \citep[e.g.,][]{baross_submarine_1985, Bad04, damer_hot_2020, russell_water_2021, WBF23, walton_cosmic_2024}. Although planet Earth is just one case study, recent works reviewed in this section have employed theoretical models to test how these consequences of the water/land trade-off might play out across exoplanetary parameter space.

\subsection{Consequences for planetary climate}


A planet's water/land ratio affects its climate through two primary mechanisms: \textit{(i)} changing planetary energy balance via surface albedo (see Section \ref{sec:climate-albedo} below); and \textit{(ii)} the role of water and erosion in the carbonate-silicate thermostat (Section \ref{sec:carbonate-silicate}). Secondarily, climate is also affected by the geographic distribution of land (e.g., supercontinents, equatorial versus high-latitude land) and by topographic relief \citep[e.g.,][]{glaser_continents_2025, SOK20, macdonald_climate_2022, del_genio_habitable_2019, way_climates_2021}.

\subsubsection{Surface albedo effects}\label{sec:climate-albedo}

Albedo --- the ratio of reflected to incident light --- is a main factor in the energy balance of a planet. All else being equal, planets with high-albedo surfaces are cooler because they reflect more incident radiation to space. For example, although ice and snow reflect $\sim$80--90\% of incoming radiation from a G-dwarf star, liquid water reflects only $\sim$5\%, absorbing the rest. Albedo is wavelength dependent, therefore both the surface composition as well as the incident stellar spectrum are important. Radiation from M-dwarf stars peaks in the near-infrared; under such a stellar spectrum ice and snow have 35--70\% lower albedo and absorb more incoming radiation compared with G-dwarf stars \citep[the albedo decrease under K-dwarf stars is 6--17\%;][]{pierrehumbert_palette_2011, shields_effect_2013}. Ice and snow, however, are always more reflective than an ocean, leading to an ice-albedo feedback effect \citep[see next paragraph;][]{shields_effect_2013}. Given relatively small land/water ratios (e.g., on Earth), ocean-absorbed heat has a major effect on climate; heat is redistributed via ocean circulation to maintain relatively moderate ocean surface temperatures across diurnal and seasonal cycles \citep{rahmstorf_ocean_2002}. Meanwhile, land reflects more energy compared with oceans --- lowering the total energy in the system --- and moreover cannot efficiently redistribute this energy spatially, leading to large diurnal and seasonal variations in temperature on continents. In this way, the evolution of continents (see Section \ref{subsec:continents}) has affected Earth's global energy balance.

Aside from vegetation cover, land surface albedo is largely affected by mineralogy. A clear distinction can be drawn between more primitive, ubiquitous basaltic crusts rich in dark mafic minerals (albedo $\sim$0.1), and the light-coloured felsic crust typical of Earth's processed continents (albedo $\sim$0.35). Indeed, Earth's bare felsic mineral surfaces are among the brightest found on the terrestrial planets in the Solar System (see Lustig-Yaeger et al., 2026, this topical collection for a discussion of exoplanet crust mineralogy and its albedo signatures). 


Perhaps the most notorious surface albedo- and climate-related phenomena is the ice-albedo feedback. Upon freezing, the albedo of water increases drastically, from $\sim$0.05 to $\sim$0.9. A decrease in temperature creates more ice, reflecting more radiation to space, further cooling temperatures --- and giving rise to a positive feedback loop \citep{thackeray_emergent_2019, zhang_review_2022}. The ice-albedo feedback is best exemplified by the `snowball Earth' glaciations of the Neoproterozoic \citep[e.g.,][among others]{fairchild_neoproterozoic_2007, chandler_climate_2000, spiegel_generalized_2010}. These glaciation periods were likely triggered, at least in part, by changes in orbital parameters (the Milankovitch cycle); the positive nature of the ice-albedo feedback led to a runaway snowball climate \citep{spiegel_generalized_2010}. Further, the presence of land has a large effect on the stability and lifetime of ice, as seen by comparing Earth's Arctic to Antarctic poles. Because oceans are able to circulate heat from warmer latitudes, polar sea ice may melt during the summer. Continual, multi-year sea ice in the Arctic is restricted to latitudes north of $\sim$80\textdegree{}N in the pre-industrial climate \citep{walsh_database_2017, comiso_bootstrap_2023}, whereas ice sheets have persisted on the Antarctic continent for millions of years \citep{yamane_exposure_2015}. The generally-decreasing albedo of ice with increasing wavelength (and hence peak radiation from increasingly smaller stars) means that the ice-albedo feedback is expected to be muted on planets orbiting M-dwarfs, and may even reverse into a stabilising feedback at high water/land ratios \citep[e.g.,][]{shields_effect_2013, rushby_effect_2019}. 

\subsubsection{Continents and the carbonate-silicate weathering thermostat}\label{sec:carbonate-silicate}

Earth's clement climate is maintained over geological timescales by the carbonate-silicate cycle \citep{Walker:1981, gislason_feedback_2008, poggevonstrandmann_lithium_2021, arnscheidt_presence_2022, brantley_how_2023}. The negative nature of this feedback maintains a climate equilibrium, where global mean temperatures have remained at $\sim$10--35\textdegree{}C for the last 450 Ma \citep{judd_485-million-year_2024}. The carbonate-silicate cycle balances greenhouse CO$_2$ concentrations via the temperature-dependent removal of CO$_2$ via reactions with calcium carbonate (CaCO$_3$). When CO$_2$ concentrations are high and the climate is warm, precipitation is stronger (increased water vapour capacity in the atmosphere) and more acidic (CO$_2$ dissolves into water to form carbonic acid). The acidic rain erodes and chemically weathers silicate rocks, releasing ions including calcium. Weathered calcium reacts with aqueous CO$_2$ (in the form of HCO$_3^-$) to produce calcium carbonate precipitate. This net process removes CO$_2$ from the atmosphere, reducing the temperature. Presumed constant over the timescales of this weathering process (hundreds of thousands of years) is the volcanic release of CO$_2$ through outgassing. Therefore, when temperatures are cooler, less CO$_2$ is removed from the atmosphere, but the same amount is introduced, increasing the atmospheric concentration and hence the surface temperature. 
Although the quantitative details of Earth's carbonate-silicate weathering thermostat are not settled, some form of negative climate feedback seems far more likely than a chance balancing of surface carbon sources and sinks \citep{coogan_regulation_2025}.

The carbonate-silicate cycle, as we understand it, may require both exposed land and water to work effectively from both sides \citep{lingam_dependence_2019,glaser_detectability_2020, krissansen2021}: 
\begin{enumerate}
    \item As argued by, e.g., \citet{glaser_detectability_2020}, seafloor weathering is less effective than subaerial continental weathering. The reasons are twofold: \textit{(i)} decreased physical erosion (rivers have much higher velocity than seafloor currents) and \textit{(ii)} increased pH in the ocean (Earth's ocean pH $\sim$8.2). Physical erosion serves to increase the surface area of the substrate, increasing weathering, and low, acidic pH serves to increase chemical weathering rates. For extremely deep oceans --- hundreds of kilometres --- high-pressure ice \citep{leger_new_2004, noack_waterrich_2016} would seem to decouple any eroded rocky material from dissolved oceanic carbon, breaking the loop.   
    \item  Volcanism and outgassing may be suppressed on planets with deep global oceans. 
    CO$_2$ would be more soluble in (i.e., exsolve less from) magma under higher pressures, so CO$_2$ outgassing may be orders of magnitude lower than if it were degassing subaerially \citep{krissansen2021}. Further, silicate melt itself could be denser than the ambient mantle at these high overburden pressures (i.e., with oceans $\gtrsim$400\,km deep), and so would not rise via buoyancy forces to the surface \citep{stolper_melt_1981, noack_waterrich_2016}. Inefficient CO$_2$ outgassing may lead to a runaway snowball state.
\end{enumerate}
Nevertheless, the kinetics of seafloor weathering remain ambiguous; several studies have argued that seafloor weathering could still be an efficient carbon sink under high-enough geothermal heat flux \citep[e.g.,][]{coogan_evidence_2013, coogan_temperature_2018, coogan_alteration_2015,krissansen-totton_constraining_2018, nakayama_runaway_2019}. More recent studies have begun to consider how effectively atmospheric CO$_2$-regulating processes might work under various extraterrestrial conditions, including crust mineralogy \citep{hakim2021} and high CO$_2$ itself \citep{graham_carbon_2024}.

\subsubsection{Planetary and numerical experiments}

In the record of Earth's supercontinent cycle, we might extricate a kind of planetary test of the relationship between changing land area and distribution on one hand, and atmospheric CO$_2$ and temperatures on the other hand. The assembly of Pangaea is associated with a marked decrease in global sea level (hundreds of metres), a decrease in CO$_2$ partial pressure (a few thousand ppm), and an ensuing icehouse climate; vice versa for Pangaea's breakup \citep[and references therein]{nance_supercontinent_2022}. Falling sea levels have been explained by supercontinents' thermal uplift and more voluminous ocean basins across the planet \citep{worsley_global_1984}. Atmospheric CO$_2$ changes follow from changes in continental weathering: not only does the falling sea level expose more land, but the colossal act of assembly means more erosion and chemical weathering, drawing down CO$_2$. Some studies have attempted to close the loop by pointing out ways global climate could itself affect mantle circulation, implying complex geological couplings through the whole planet \citep{foley_whole_2016, jellinek_ice_2020}.


Absent any other geological record, an important tool for testing how water/land ratios and configurations influence planetary climate is a general circulation model (GCM). GCMs become especially useful for exploring the exotic climates of exoplanets with wholly non Earth-like orbital parameters such as 1:1 spin-orbit resonance (i.e., tidally locked). The climate of tidally locked planets are of interest as most habitable zone exoplanets in M-dwarf systems are likely tidally locked \citep{barnes_tidal_2017}. Pioneering 3D GCM studies have demonstrated possible `eyeball' climate states on these tidally locked, habitable zone exoplanets, where liquid water would be stable directly under the star (the substellar point), given surface oceans at least $\sim$10\% the mass of modern Earth's \citep[e.g.,][]{
pierrehumbert_palette_2011, yang_water_2014, turbet_habitability_2016}. Drier planets may see a longitudinal ring of temperate conditions, along the permanent `evening' terminator, due to reduced heat transport in the atmosphere \citep{lobo_terminator_2023}. If realistic ocean heat transport (i.e., dynamic ocean) is included in the GCM, then the `eyeball' appears to stretch into a `lobster' \citep[e.g.,][]{hu_role_2014, yang_water_2014, delgenio_albedos_2019}. Later works have begun introducing new continent configurations to the experiment \citep[e.g.,][]{lewis_influence_2018, SOK20, ZLL21, macdonald_climate_2022, way_trappist1_2025} --- in particular, featuring variations on the theorised scenario of such planets having single large landmasses/basins continuously re-oriented to the substellar point, via true polar wander \citep{leconte_continuous_2018}, like the Moon \citep{zuber_shape_1994}. These experiments have concluded that increasing (dayside) water/land ratio decreases global mean temperature by $\sim$20\,K, mostly via increasing the planetary albedo, although the substellar point locally heats up. In fact, the dayside may become a desert, reducing the evaporation of water vapour from the ocean into the atmosphere (and thus the impact of this water vapour on longwave radiation). Similar trends have also been shown in simpler energy-balance models \citep{rushby_effect_2019, honing_land_2023}. These consequences of land/water surface albedo on global mean temperature persist across different stellar spectral classes, from M- to F- dwarfs, despite the changes to absolute albedos of land and water \citep{rushby_effect_2019}. However, in F-G-K systems, continental configurations may be less important, since habitable zone exoplanets in these systems are likely not tidally locked (i.e., fast rotating) and do not have permanent day/night sides \citep{glaser_continents_2025}. Overall, the complexities of atmosphere and ocean heat transport, and the sensitivity of surface albedo to host star spectrum, mean that when investigating a specific planet in practice, a number of targeted simulations should be carried out to delineate a range of possible climates.

\subsection{Consequences for life}

The biomass on modern Earth is not evenly distributed between land and  oceans \citep{bar-on_biomass_2018}. While terrestrial biomass comprises 470 Gt carbon of mostly plant life (autotrophic, although only one third is active), marine biomass totals 6 Gt carbon of mostly heterotrophic life. Motivated in part by this simple observation, we can ask how and why the planetary land fraction would affect a propensity for biological productivity. Tightly linked with both land fraction and life is climate: productivity in the seas depends on sea surface temperature, for example; and such effects have been studied through coupling climate and biogeochemical models \citep{lerner_obliquity_2025}. Here we focus on direct trade-offs of water versus land surfaces on life.


\subsubsection{Net primary productivity}

Comparing markers of biological productivity between Earth's land and oceans provides a clue to the hypothetical consequences of water/land ratios on life. Net primary productivity (NPP) encapsulates the net amount of organic carbon produced through photosynthesis (expressed in units of kg C yr$^{-1}$). It regulates many facets of the biosphere \citep{SB13}. In the case of Earth, the terrestrial and marine NPP are roughly equal to each other \citep{Field:1998,2011GCBio..17.3161I}. The lesser areal proportion of land means that NPP per unit area is greater on land than in the oceans. As mentioned above, the modern land biomass ultimately dominates the marine biomass by about two orders of magnitude \citep{bar-on_biomass_2018} --- though section \ref{sec:life-time} will note how this fact does not hold for all of Earth history. The large difference in biomass and smaller difference in NPP seem related to how biomass is distributed across trophic levels: marine life is overabundant in consumers, which build biomass inefficiently compared to primary (photosynthesising) producers; moreover, land-based producers tends to have a longer turnover time \citep{hatton_predatorprey_2015, bar-on_biomass_2018}. If the physical mechanisms behind marine life's inverted food pyramid \citep[see][]{burgess_scale_2018} would operate elsewhere, then other marine biospheres may be similarly less productive than their land equivalents.

On Earth, the maximum NPP across more than half of all land area is limited by access to liquid water resources \citep{CR98}. On land planets with limited water inventories, it is conceivable that the surface hydrological cycle would serve as a bottleneck on NPP, albeit not the only such control (e.g., temperature is another key determinant). Meanwhile, Earth's marine biosphere is limited by nutrient availability: chiefly phosphorus (P) in the form of phosphates, delivered by way of continental weathering \citep{SB13}. Whilst there has been debate about what nutrients serve as the ultimate bottleneck on NPP, many works support phosphorus in not just the Phanerozoic, but also the Archean and the Proterozoic eons \citep{TT99,LS18,HKH20,ZMP24}. Several works have suggested that NPP on ocean worlds could be stymied by dissolved P availability because of diminished continental weathering \citep{lingam_dependence_2019,LL21,glaser_detectability_2020,OJA20}.  A simple model developed by \citet{lingam_dependence_2019} in this context suggested that the NPP and biomass of both (water-limited) land and (nutrient-limited) ocean planets are potentially orders of magnitude lower than on Earth. 

However, alternative pathways for nutrient supply to marine biospheres may be found in oceanic crust weathering and/or serpentinisation \citep{SRI21,POF22}. Besides phosphorus, other candidates for the limiting nutrient are nitrogen \citep[as nitrates;][]{TT99,BMA17}, iron \citep{BMA17, wade_temporal_2021}, or trace metals such as molybdenum \citep{AK02}. The factors potentially modulating the availability of these alternative limiting nutrients on a given planet range from volcanic outgassing rates and hydrothermal activity to atmospheric redox states \citep[e.g.,][]{lingam_dependence_2019}. In any case, we highlight that the same single ingredient need not limit biological productivity for the entire history of life on a planet. \citet{wade_temporal_2021} propose that it was the \textit{decrease} in environmental Fe$^{3+}$ availability after the Great Oxygenation Event that eventually led to the evolution of multicellular life, able to efficiently recycle iron within one organism.


This discussion has painted a heuristic picture of how NPP could be sensitive to a planet's water/land ratio, but many other aspects are potentially important. For instance, provided that nutrient availability is not a constraint, the average ocean depth plays a key role in governing NPP. Most biological productivity in Earth's oceans occurs in the top $\sim$200 m \citep{SH12}, stemming from a combination of nutrient mixing and upwelling driven by wind and tidal forces \citep{RC08}, as well as sufficient photon fluxes for photosynthesis \citep{SG06}. On planets that experience strong tidal forces --- either from the host star or a large moon --- it is conceivable that enhanced tidal mixing might help boost NPP \citep{LL18,OJA20}, if all other factors are held fixed.

\paragraph{Atmospheric oxygen} 

On Earth, the NPP also shares close links with the production of molecular oxygen (O$_2$) because the predominant pathway for biosynthesis is oxygenic photosynthesis. As a result, worlds with low NPP would be expected to have limited fluxes of O$_2$ production. Hence the endmember water/land ratios of 0 or 1 may be associated with low atmospheric O$_2$ levels even on inhabited planets; this qualitative reasoning is broadly supported by simple quantitative models \citep{lingam_dependence_2019,LL21,WT21}. We add the caveat that atmospheric O$_2$ levels are governed not only by O$_2$ sources (oxygenic photosynthesis) but also by sinks, determined by weakly bound variables such as carbon burial efficiency and the flux of reducing gases.

If the prediction that all-land or all-ocean planets have lower atmospheric O$_2$ is valid, further evolutionary consequences become apparent. It has been proposed that a relatively high concentration of atmospheric oxygen is a prerequisite for the emergence and/or diversification of complex multicellularity, given the bioenergetic advantages of aerobic respiration \citep{Knoll85,CGZ05}. Likewise, O$_2$ levels comparable to modern Earth might be valuable or even necessary for technological intelligence \citep{LBM23,BF24,stern_2024}.

\subsubsection{Life in deep time}\label{sec:life-time}

The origin and diversification of life on Earth both seem to be inseparable from its patterns of land and ocean coverage. For each mechanism to which we may owe the beginnings of life on early Earth, the water/land ratio seems to matter \citep[e.g.,][and references therein]{scherf_eta-earth_2024}. Ocean planets are unable to facilitate the wet/dry cycles critical to the warm little ponds origin of life hypothesis; land planets lack the redox and elemental gradients key to the alkaline submarine hydrothermal vent origin of life hypothesis. Either way, there will not be a clear sense of change of biological diversity with increasing water/land ratio, if only for the fact that evolution has a nonlinear history. The fossil record suggests that the biosphere occupied land only from the Phanerozoic onwards \citep[e.g.,][]{awramik_evolution_2007}: land life owes itself to earlier evolution in the oceans.

Planetary land fractions and their evolution in time (e.g., supercontinent cycles) would then go on to influence how (complex) life might evolve. By drawing on the major evolutionary transitions that unfolded on Earth \citep{SMS95} --- some of which definitely occurred on land, with others unfolding in water --- \citet{lingam_dependence_2019} suggested that the relative odds of these events occurring on either pure-land or pure-ocean worlds is suppressed by orders of magnitude relative to Earth. The optimal chances for major evolutionary transitions to complex life could be found on worlds with roughly equal surface fractions of landmasses and oceans. 

In their discussion of so-called `superhabitable' worlds, \citet{heller_superhabitable_2014} raised the point that it was not just the total area of landmasses or oceans that would be crucial, but also their distribution. Indeed, empirical trends documented on Earth show that coastal benthic environments are more diverse than open water\citep{gray1997marine}. This theme was elaborated by \citet[Chapter 5.6]{LL21} by drawing on established models from theoretical ecology \citep{MLR95,MM07,harte2011maximum}, chiefly the principle of maximum entropy and the species-area relationship, to outline how land distributed over smaller regions could conceivably yield greater biodiversity than a single supercontinent having the same area.

Moreover, biodiversity is observed to correlate with tropical regions that may change with time due to continental drift. As Alexander von Humboldt (1769--1859) already knew, mountain belts, particularly the Andes, are exceedingly rich in biodiversity \citep{rahbek2019humboldt}. In a similar vein, marine animal diversity has been observed to increase during the breakup of supercontinents \citep{ZFP17}, possibly stemming from changes in the areas of continental, coastal, and deep water environments. The converse might arise during the assembly phase. This apparent correlation between speciation and the supercontinent cycle raises the possibility that worlds upon which this cycle operates on different timescales --- or essentially not at all (e.g., on ocean worlds) --- might display speciation dynamics distinct from Earth. 



\section{Observing water/land ratios and evaluating the presence of life}
\label{sec:EPSI}

Seeing oceans, continents, quasi-static weather patterns, atmospheric and surface biosignatures, and even artificial structures on exoplanets would all be clues for detecting and characterising life outside the Solar System. In fact, as discussed in the previous sections, the water/land ratio itself can inform us about a planet's climate and possible geodynamics, as well as the feasibility of extant life-as-we-know-it and its diversity.

This section first reviews recent modelling efforts to extract planetary surface coverage information from (hypothetical) observations of rocky exoplanets and their validation from Earth observations. Then, it elaborates on future astronomical observational capabilities to collect necessary data and obtain the first maps of rocky exoplanets in habitable zones.

\subsection{Mapping surfaces of temperate rocky exoplanets and evaluating their water/land ratio}
\label{sec:EPSI_maps}


Time-series measurements of the light reflected from an exoplanet, recorded over the course of the planet's axial rotation and orbital motion --- that is, light-curves --- contain information on structures on the planetary surface, as well as in the atmosphere and near-planetary space. Such one-dimensional (1D) time-series of light-curves can provide two-dimensional (2D) time-averaged albedo maps with structures resolved in both longitude and latitude. For more than hundred years, a deconvolution of the longitudinal and latitudinal information has been proven to be feasible, thanks to a favourable combination of the axial planet rotation --- revealing different hemispheres --- and different illumination and reflection angles due to planet orbital motion, especially when the axial and orbital inclination (obliquity) directions are misaligned. 


Model solutions obtaining albedo maps, along with the spin-orbital parameters, from reflected flux or polarization light-curves have been previously demonstrated for angularly unresolved planets, moons and asteroids in the Solar System, for and various types of exoplanets, from hot Jupiters to Earth-like \citep[e.g.,][]{russel1906, guthnick1906, morrison1975, Buie1997, kaasalainen1992, carbognani2012, cowan_alien_2009, cowan_odd_2017, fluri_orbital_2010, fujii_colors_2010, kawahara_mapping_2011, fujii_mapping_2012, schwartz_inferring_2016, lustig-yaeger_detecting_2018, farr_exocartographer_2018, berdyugina_surface_2019, fan_earth_2019, aizawa_global_2020, kawahara_global_2020, kawahara_bayesian_2020, asensioramos_planet_2021, gu_earth_2021, gu_earth_2022, kuwata_global_2022, teinturier_mapping_2022}. These works differ by assumptions and by numerical approaches to forward modelling and deconvolution (inversion) of light-curves, which affect properties of inferred `best' solutions and their uncertainties; that is, how one starts from simple 1D back-projection `maps' to 2D principle-component maps and pixel-wise inversions that result in multi-colour 2D maps. Assumptions have differed about, for example, star-planet geometries and compositions of the planetary surface and atmosphere. Methods for mapping surfaces of exoplanets (exo-cartography) have been previously reviewed by \citet{berdyugina_rev_2019} and \citet{cowan_mapping_2020}.


In the context of future capabilities for reflected light detections from rocky exoplanets, many of the above-cited works have validated the principles of exo-cartography using Earth as an available test-case of such exoplanets. Since \citeauthor{sagan_search_1993}'s analysis of Earth observations by the Galileo spacecraft \citep{sagan_search_1993}, Earth-as-an-exoplanet data have been obtained from the EPOXI mission \citep[e.g.,][]{cowan_alien_2009, cowan_rotational_2011, robinson_earth_2011, fujii_colors_2011}, 
the NASA Earth Observatory \citep[NEO; e.g.,][]{berdyugina_surface_2019}, the Moderate resolution Imaging Spectroradiometer \citep[MODIS; e.g.,][]{kawahara_mapping_2011, mettler_earth_2020, kelkar_earth_2025}, the Deep Space Climate Observatory \citep[DSCOVR; e.g.,][]{jiang_using_2018, fan_earth_2019, aizawa_global_2020, gu_earth_2021, gu_earth_2022, kuwata_global_2022}, and the Lunar Crater Observing and Sensing Satellite \citep[LCROSS; e.g.,][]{robinson_detection_2014}. 

\begin{figure}
    \centering
    \includegraphics[width=0.51\linewidth]{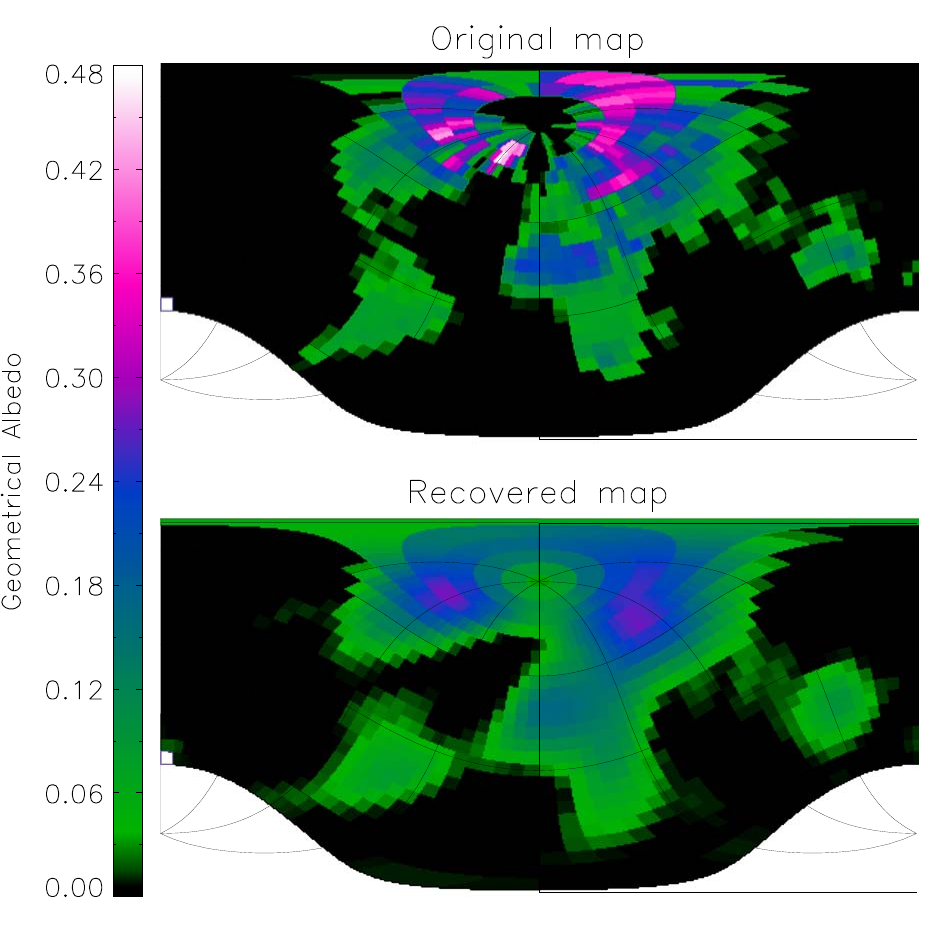}
    \includegraphics[width=0.4\linewidth]{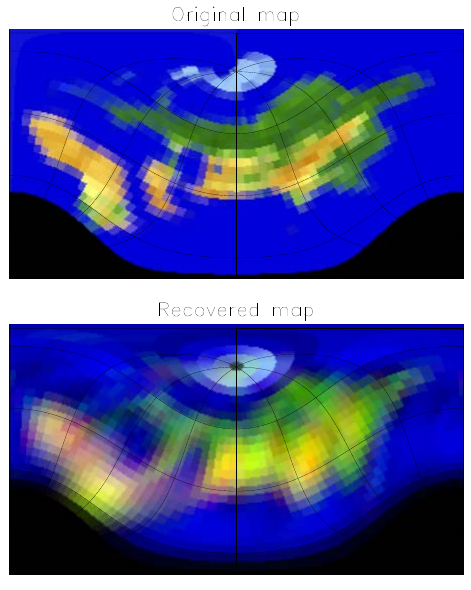}
    \caption{Model cases for a distant Earth-like exoplanet indirectly imaged using the ExoPlanet Surface Imaging technique \citep[EPSI;][]{berdyugina_surface_2019}. \emph{Left:} The original map is based on NASA Earth Observatory albedo data from March 2003. The recovered map is obtained by inversion of a reflected light curve synthesized using the original map, assuming a signal-to-noise ratio of 200 and a total number of measurements of 3000. \emph{Right:} The original map is based on a composite (RGB) Earth image (similar to that in Fig.~\ref{fig:blue marble}) with parts of the terrestrial continents arranged in a plausible pattern. The recovered map is inferred from four broad-band (optical to NIR) reflected light curves. Continents and their spectral features are well recovered \citep[for details, see][]{berdyugina_surface_2019}.}
    \label{fig:EPSI_maps}
\end{figure}


Since dry land is typically of a higher average albedo compared with the very low albedo of water oceans (Section \ref{sec:climate-albedo}), and since (Earth-like) land vegetation has characteristic spectral and polarised reflectance \citep{sagan_search_1993, seager_vegetations_2005, kiang_spectral_2007, kiang_spectral_2007a, berdyugina2016, arp_quieting_2020}, inversions that are robust to measurement errors can reveal contours of continents, even for sparsely-measured and noisy light-curves (Fig.~\ref{fig:EPSI_maps}). Thus the water/land ratio is the first, immediate outcome of reflected light-curve inversions, if land masses and oceans are both present, and if the entire planetary surface is observed and visible at a given wavelength. For exoplanets that are tidally-locked at different spin-orbital resonances (e.g., Proxima b), extracting a water/land ratio on an observable part of the planetary surface was demonstrated by \citet{berdyugina_surface_2019}: the continents can be well mapped if present in the star-facing planetary hemisphere.

A common issue remaining in all map inversion techniques lies in how to account for time-variable clouds (weather) and surface features (seasons). For example, using MODIS data, \citet{kawahara_mapping_2011} showed that light-curve inversions can recover an approximate cloud distribution, whereas inversion of light-curve \textit{differences} between near-infrared (0.8--0.9\,$\mu$m) and blue (0.4--0.5\,$\mu$m) bands roughly recovers the water/land distribution, except for at high-latitude regions persistently covered with clouds and snow. This study considered the case of a partially cloudy mock Earth-twin on a face-on orbit. From time-series of NEO albedo data, \citet{berdyugina_surface_2019} inferred realistic surface albedo maps with varying cloud cover, as well as time-dependent partial maps showing seasonal changes in surface vegetation and ice cover. In other works \citep{fan_earth_2019, aizawa_global_2020, kuwata_global_2022}, two principle components ascribed to land coverage and to `surface-independent' cloud coverage were extracted from DSCOVR data; Earth's 2D water/land distribution was roughly reproduced. \citet{teinturier_mapping_2022} tested a cloud-removal scheme using DSCOVR observations, but found it difficult due to the necessarily full-phase data. Motivated by the need to account for cloud and surface variability, further advancement in numerical approaches, such as Bayesian dynamic mapping \citep{kawahara_bayesian_2020} and regularization learned from mock surfaces \citep{asensioramos_planet_2021}, has delivered model reconstructions of large-scale evolving patterns. Interpretation of such high and low albedo patterns on real exoplanets with unknown geography and weather varieties will nonetheless be a challenge.

How much detail can be inverted from an exoplanet's reflected light? For sufficiently high signal-to-noise ratio (S/N$\sim$20 or more) and number of measurements (200 or more), the ExoPlanet Surface Imaging (EPSI) technique of \citet{berdyugina_surface_2019} could detect surface features down to about 2\%\ of the total exoplanetary surface area, such as Sahara desert (1.8\%) and Australia (1.5\%; see Fig.~\ref{fig:EPSI_maps}). \citet{aizawa_global_2020} also showed that low-noise and frequent measurements of exoplanets comparable with the disk-integrated DSCOVR Earth data are needed to produce reliable results. \citet{kawahara_mapping_2011}, employing simulated Earth-twin data also with S/N=20, were able to infer large-scale cloud cover and surface structures. 

Even more detail would be possible with multi-wavelength light-curve measurements (optical to near-infrared bands). These observations could provide `photographic' views of distant exoplanets, analogous to Earth images (Fig.~\ref{fig:blue marble}) but certainly at a lower spatial resolution. In \citet{berdyugina_surface_2019}, this ability was demonstrated for Earth, other Solar System planets and moons, and simulated exoplanets with Earth-like life and various artificial structures. In particular, low-resolution reflected spectra consisting of only four broad-band measurements sufficed to distinguish spatially resolved patches of various surface compositions, such sand and vegetation. Similarly, \citet{kawahara_global_2020} demonstrated performing a spectral retrieval (ten photometric bands) of several surface components for a composite colour surface map of Earth as an exoplanet. Hence such multi-wavelength images of exoplanets would in principle provide information on the surface composition, geology, geodynamics, climate, and possibly even the presence of life.

The geographic location of land masses with respect to the equator and poles can be also well documented in such maps ( Fig.~\ref{fig:EPSI_maps}). This result is achieved via a simultaneous retrieval of the map and spatial orientation of the spin and orbital axes of the planet in several currently employed methods. In particular, the obliquity of the spin axis with respect to the orbit may inform us about planet's possible seasons and climate zones, similar to those on Earth. Errors on the spin-orbital parameters can vary depending on the quality of data. However, an additional constraint on planetary obliquity with respect to the observer would be provided by the detection of a polar ice cap \citep{berdyugina_surface_2019}.

A completely different approach to mapping exoplanetary surfaces is based on the remarkable optical properties of the solar gravitational lens (SGL), whose focal line begins at $\sim$540 AU from the Sun. The SGL can be employed to focus light from a faint, distant source along that line by bending photon trajectories. Advantages include major brightness amplification by a factor of $\sim10^{11}$ at the wavelength of 1\,$\mu$m and extreme angular resolution of $\sim10^{-10}$ arcsec within a narrow field of view \citep{TuryshevToth2017}. To make use of this opportunity, it was proposed to position a telescope beyond $\sim$540 AU and use the SGL to magnify light from distant objects on the opposite side of the Sun; for example, from Proxima b \citep{Eshleman1979, Maccone2009, TuryshevToth2017, Turyshev2018, TuryshevToth2022}. Technical feasibility studies for such an interstellar space telescope are being carried out \citep[e.g.,][]{TuryshevToth2023}. The SGL telescope may deliver unprecedented high-resolution imaging and spectroscopy of terrestrial exoplanets within $\sim$30--100 parsec from the Sun, as a possible follow-up mission for the most interesting target. 

\subsection{Future facilities for mapping surfaces of potentially habitable exoplanets}
\label{sec:EPSI_obs}

Detecting reflected light from rocky exoplanets in habitable zones of cool main-sequence stars --- i.e., G-, K- and M-type dwarfs --- is currently beyond capabilities of optical astronomical facilities on the ground and in space. The observations are difficult because the reflected light from the planet is many orders of the magnitude dimmer than the direct light from the star. A detection of these planets' reflected light at high S/N requires an imaging contrast of at least 10$^{-8}$ to 10$^{-10}$, for planets around early M-dwarfs to early G-dwarfs, respectively. Achieving such a high imaging contrast requires special optical designs and data calibration techniques: stellar coronography and high angular resolution, for example. Also required to reduce the stellar light contamination in the reflected light is a large angular separation between the host star and its planet.

These optical requirements can be in principle achieved with extremely large-aperture ground-based telescopes, such as the Extremely Large Telescope (ELT); the Giant Magellan Telescope (GMT), or the Thirty Metre Telescope (TMT). Considering the imaging contrast achieved so far with existing large ground-based telescopes (e.g., the Very Large Telescope, VLT; Gemini), which is about 10$^{-6}$ to 10$^{-7}$, the coming generation of extremely large telescopes may be suitable for indirect light-curve imaging of rocky exoplanets in the habitable zones of late (coolest) M-dwarfs and early (hottest) brown dwarfs, if these observatories achieve a similar contrast \citep[e.g., for ELT, current efforts are focused in the infrared,][]{Feldt_ELT_2024}. For instance, observations of Proxima b with a 30--40\,m aperture telescope may deliver reflected light measurements with S/N of 10 to 100 in the photometric BVRI-bands (i.e., blue-to-infrared), respectively, for an exposure time of only one hour, when assuming reasonable planetary parameters and telescope efficiency \citep{berdyugina_surface_2019}. A 20m-class telescope may be able to achieve an S/N of 5--20 in the BVRI-bands in one hour. In fact, several dozens of habitable zone exoplanets around M-K-G-dwarfs within the solar neighbourhood (limited by the stellar magnitude; brighter than $V$ = 13 mag) can be detected in reflected light with S/N$\ge$5, using a 25-m telescope within eight hours of total exposure time. Several high-contrast telescope designs with apertures up to 100\,m have been proposed for the purpose of rocky exoplanet mapping: for example, the 25-m interferometric telescope Exo-Life Finder \citep[ELF;][]{Kuhn_ELF_2018,Berdyugina_ELF_2018}, and the 100-m OWL-Moon telescope on the Moon \citep{Schneider2022}, among others. 

Most recently, design studies have been initiated for the NASA space-based Habitable Worlds Observatory (HWO) with aperture up to 8\,m (see Lagage et al., 2026, this topical collection for a detailed discussion). The design of HWO aims to achieve at least the 10$^{-10}$ contrast at angular separations corresponding to the habitable zones of nearby solar-type stars, using stellar coronagraphy and multiple differential techniques including spectroscopy and polarimetry. Achieving such a high imaging contrast will open a unique opportunity for obtaining multi-spectral albedo maps of temperate rocky exoplanets and detecting photosynthetic life in the solar neighbourhood \citep{Berdyugina2025}. HWO science case studies and scientific requirements provide estimates of possible number of detections and exposure times necessary to achieve these and other scientific goals of the HWO mission.

\section{Conclusion: On the prevalence of water versus land planets}
    


This review has demonstrated how the water/land ratio on a temperate planet is an intricate outcome of internal mantle dynamics. In turn, oceans trade off with land surfaces to exert a grand influence on climates and biospheres. One might legitimately ask whether Earth's balanced water/land ratio is a coincidence, which could possibly augment the attributes that have been posited to make Earth a rare planet \citep[e.g.,][see also Spohn et al., 2024, this topical collection]{scherf_eta-earth_2024}, or whether the ratio is a natural outcome of planetary processes. There is no definitive answer to this question yet, but we have the tools to approach it. We consider three ingredients that were previously discussed, noting that there is no evidence for these ingredients requiring plate tectonics in order to perform their functional purpose, although recycling in a mobile-lid mode may enable ingredient 3 especially.

First, to keep dry land above sea level, the planet must support substantial surface topography. Conveniently, there are many ways to create this topography. Although Earth's elevation profile is undoubtedly complex, the product of billions of years of plate movement and mountain-building, large-scale flows in the mantle will nonetheless express themselves as long-wavelength dynamic topography at the surface. All convecting rocky planets thus have some intrinsic topography. 

Less certain beyond Earth, but potentially dramatic in shaping ocean basins, is the isostatic flotation of low-density felsic `continents' above the denser mafic crust that progenates them. The formation of continental crust is related to the water cycle between the mantle and the surface, at least in the context of modern plate tectonics \citep[e.g.,][]{honing_land_2023}, but may also matter for tectonic modes that may have dominated Earth during the Hadean and Archean. Unfortunately, the rock record is sparse for these periods and the early volume of continental crust (compare Fig. \ref{fig:Crust Age}) and the mechanism(s) of their formation and growth are still uncertain \citep[e.g.,][]{cawood_secular_2022, chowdhury_continental_2024, rey_archean_2024}.

At finer scales, a number of other surface processes are known to `roughen' one-plate planets in the Solar System: lava doming, impact cratering, chemical dissolution of the crust, or any means of tectonic extension or compression, for example. Regardless of how topography is generated, however, gravity will ultimately cap its maximum prominence due to finite lithospheric strength, such that Earth-size planets likely do not support mountains surpassing 10 km \citep{melosh_strength_2011, guimond_blue_2022}. Nevertheless, the large scope for planetary topography at all scales suggests that we can nevertheless expect environments where liquid water, if present, can at least pond if not form deep oceans.

\begin{figure}
    \centering
    \includegraphics[width=1\linewidth]{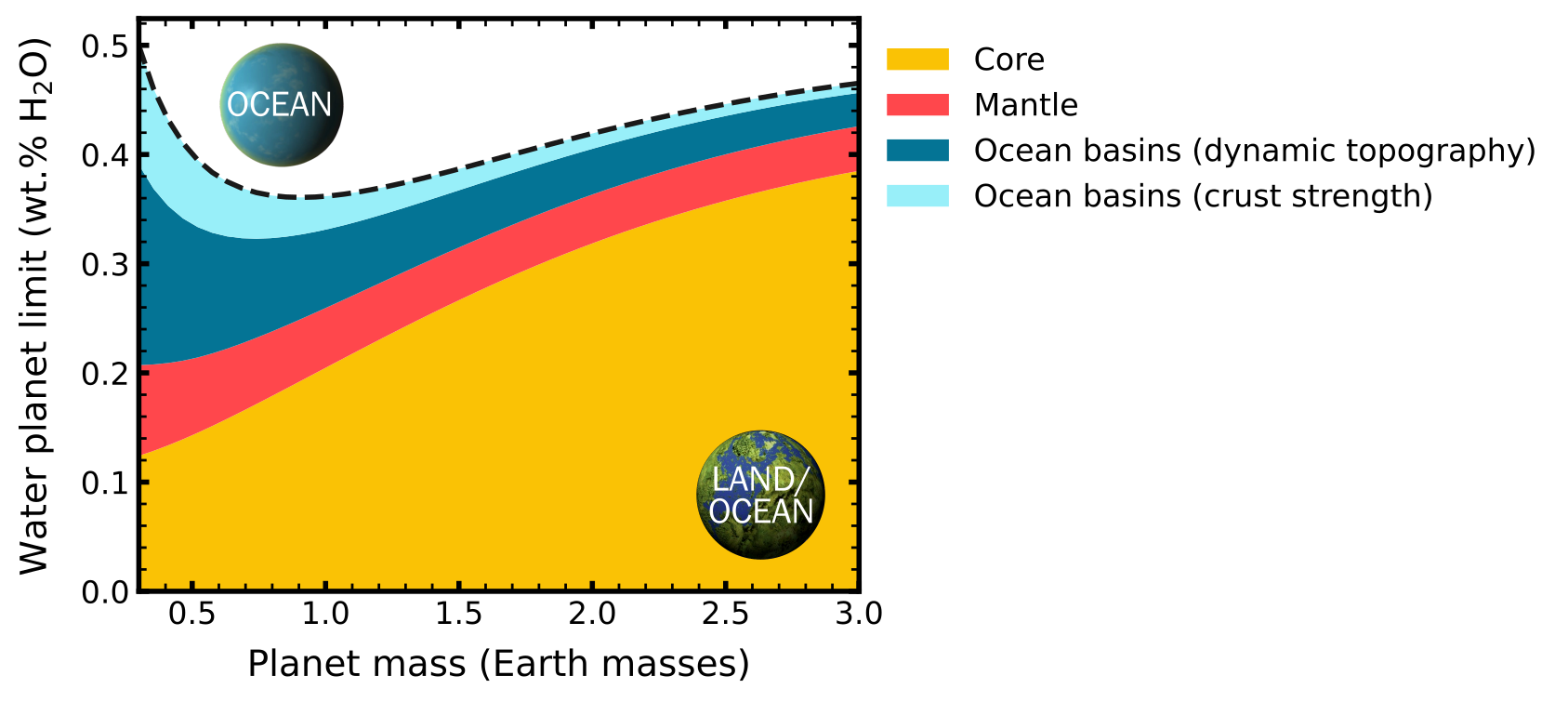}
    \caption{The minimum bulk water mass fraction required to completely cover a planet in oceans. Coloured swaths represent the contribution to this water capacity of each of the planet's main water reservoirs, while the dashed line shows the sum of all reservoirs. Water mass fractions above the dashed line would likely make a water planet with no dry land. Ocean basin capacities assume purely-dynamic topography (dark blue contribution), plus the possibility of additional topography (light blue contribution) capped at where gravity exceeds a crust strength of 100 MPa \citep{guimond_blue_2022}. Mantle water content at water saturation (red contribution) is the median from \citet{guimond_mantle_2023}, assuming a solidified mantle with water stored only in nominally anhydrous minerals at a conservative potential temperature of 1600 K. Core water content (yellow contribution) is solved iteratively using the parameterisation for metal-silicate partitioning of H$_2$O from \citet{luo_majority_2024} given the mantle plus surface water mass above (i.e., assuming no loss to space), which depends on the water concentration in the core and on pressure (taken to be half the core-mantle boundary pressure), and is qualitatively similar to \citet{li_earths_2020}. Core mass fraction is assumed to be 32.5\%.}
    \label{fig:wmf}
\end{figure}

Second, it is essential that the bulk planet water inventory is sufficiently small to preclude oceans tens to hundreds of kilometres deep. For an estimate of how the upper limit on this water mass fraction scales with planet mass, we add ocean basin capacity scalings for topography \citep{guimond_blue_2022} on top of water-saturated solid mantles \citep{guimond_mantle_2023} and pressure-dependent H$_2$O partitioning into the core \citep{luo_majority_2024}. The resulting `water planet limit' in terms of planet mass fraction increases with planet mass, but stays safely below 1 wt.\% (Fig. \ref{fig:wmf}). 
The largest uncertainties on an appropriate bulk water mass fraction limit lie in the capacities of the metallic-iron core and the perovskite lower mantle to sequester water. These saturation limits depend on the total core and mantle masses, as well as on their temperatures. (Unknown) water hidden in the core is nonetheless less relevant for water on the surface. We conclude that on temperate planets with solidified mantles, bulk water contents of more than half a percent water by mass may prohibit dry land, based on the limited capacities of these planets' interiors and ocean basins to contain water,\footnote{Caveats abound to this simplified analysis; for example, very large masses of primordial water may enter the core during differentiation, and the associated large masses in the magma ocean are subsequently lost to space.} in agreement with \citet{scherf_eta-earth_2024}. Accretion within the snowline \citep[e.g.,][]{tian_water_2015, lichtenberg_bifurcation_2021} could be important in keeping water mass fractions below this limit. 

Considering such dry yet not dessicated planets, we can compare ocean basin capacities with water outgassing rates to speculate on the prevalence of ocean and desert worlds conservatively. If dynamic topography represents the minimum `intrinsic' ocean basin capacity, then absent any other topography an Earth-mass planet would flood under $\mathcal{O}\left(0.1\right)$ ocean masses \citep{guimond_blue_2022}. Meanwhile, volcanic outgassing rates of water can be up to $\mathcal{O}\left(0.1\right)$ ocean masses per Gyr on oxidised stagnant lid planets (see Table \ref{tab:outgassing}), but also as low as $\mathcal{O}\left(0.001\right)$ ocean masses per Gyr under certain conditions. Peak singular episodes of water degassing seem to occur in the late stages of magma ocean crystallisation \citep[e.g.,][]{hamano_emergence_2013, salvador2017}. Unless a temperate rocky exoplanet has retained a bulk silicate water mass fraction of only $\textless$10$^{-4}$ from formation or its volcanic outgassing is extremely inefficient, and assuming steam can condense, it is plausible that its ocean basins will be filled by the time we observe it. 

The comparison above highlights the need for the third ingredient: the existence of feedback mechanisms that regulate sea level. Water exchange between surface and interior may be governed by the pressure dependence of water solubility in magma, which regulates degassing at the seafloor \citep[e.g.,][]{rupke_geological_2024}. 
Rates of outgassing and ingassing are also subject to the effect of water on decreasing both the viscosity and the solidus temperature of mantle rock \citep[e.g.,][]{sandu_effects_2011, seales_deep_2020}. Within the paradigm of plate tectonics, it is noteworthy that both the primary loci of water subduction and degassing are arc-related subduction zones, such that stable equilibrium states exist when the length of subduction zones is either minimised or maximised \citep{honing_land_2023}, Earth's balanced land/water ratio being close to the latter.

Geochemical data suggest that the early continents were flooded, with low relief 
\citep{rey_archean_2024}. Planet cooling and ocean crust subsidence may then have led to the rise of the continents. Plate tectonics, operating since at least the Proterozoic, may have stabilized the surface water inventory, as evidenced by the long-term constancy of the freeboard. In any case, too little is known about the early Earth to suggest a simple or universal mechanism that would result in a balanced water/land distribution. 

Clues to the possibilities raised throughout this chapter may ultimately be found on rocky exoplanets. A diversity of various water/land scenarios can be unveiled through indirect imaging of planetary surfaces using advanced multi-colour inversions of reflected light-curves \citep[e.g.,][]{berdyugina_surface_2019, kawahara_global_2020}, if suitable data is delivered by the planned large-aperture and high-contrast telescopes on the ground and in space, such as the ELT and HWO.
By determining water/land ratios on rocky exoplanets, 
we will start to see whether `intermediate', Earth-like ratios exist on other worlds, hence whether water is safely delivered to and retained on these planets, and whether their deep water cycles exhibit the self-regulation to keep such a balance. 

\backmatter



\bmhead{Acknowledgements}
CMG is supported by the UK Science and Technology Facilities Council [grant number ST/W000903/1]. PAC acknowledges support from Australian Research Council grant FL160100168. Two anonymous reviewers have provided feedback which has improved this manuscript. CMG thanks Michael Way, Tad Komacek, Rob Spaargaren, Josh Krissansen-Totton, Harrison Nicholls, and Rob Law for additional feedback and clarification.

\section*{Declarations}
The authors declare no competing interests.



\bibliography{references}

\end{document}